\documentclass[useAMS,usenatbib]{mn2e}
\usepackage{psfig, epsf, epsfig}
\newcommand{\oiii}{\hbox{[O\,{\sc iii}]}}
\newcommand{\nii}{\hbox{[N\,{\sc ii}]}}
\title[Coexistence of starbursts and AGN]{Numerical simulations
on the relative importance of
starbursts and AGN in ultra-luminous infrared galaxies}
\author[Kenji Bekki, Yashuhiro Shioya, \& 
Matthew Whiting]{Kenji Bekki$^{1}$\thanks{E-mail:
bekki@bat.phys.unsw.edu.au},
Yasuhiro  Shioya$^{2}$,
and Matthew Whiting$^{3}$\\
$^{1}$School of Physics, University of New South Wales,
              Sydney 2052, NSW, Australia\\
$^{2}$Physics Department, Graduate School of Science and Engineering, 
Ehime University, 
2-5 Bunkyo-cho, Matsuyama, Ehime 790-8577, Japan
\\
$^{3}$Australia Telescope National Facility, CSIRO, P.O. Box 76,
              Epping NSW 1710, Australia\\}
\begin{document}

\date{Accepted, Received 2005 February 20; in original form }

\pagerange{\pageref{firstpage}--\pageref{lastpage}} \pubyear{2005}

\maketitle

\label{firstpage}

\begin{abstract}

We investigate the relative importance of starbursts and AGN
in nuclear activities of ultra-luminous infrared galaxies
(ULIRGs) based on chemodynamical simulations combined with
spectrophotometric synthesis codes.
We numerically investigate both the gas accretion rates 
(${\dot{m}}_{\rm acc}$) onto super massive
black holes (SMBHs) and the star formation rates 
(${\dot{m}}_{\rm sf}$) in ULIRGs formed
by gas-rich galaxy mergers and thereby discuss what powers ULIRGs.
Our principal results, which can be tested against observations,
are as follows.

(1) ULIRGs powered by AGN can be formed by major merging between
luminous, gas-rich disk galaxies with prominent bulges containing
SMBHs, owing to the efficient gas fuelling 
(${\dot{m}}_{\rm acc} > 1 {\rm M}_{\odot}$ yr$^{-1}$) of the SMBHs.  
AGN in these ULIRGs can be  
surrounded by compact poststarburst stellar populations
(e.g., A-type stars).

(2) ULIRGs powered by starbursts 
with ${\dot{m}}_{\rm sf} \sim 100 {\rm M}_{\odot}$ yr$^{-1}$
can be formed by 
merging between gas-rich disk galaxies with small bulges
having the bulge-to-disk-ratio ($f_{\rm b}$) as small as 0.1.   

(3) The relative importance of starbursts and AGN can depend
on physical properties of merger progenitor disks, such as
$f_{\rm b}$, gas mass fraction, and total masses.
For example, more massive  galaxy mergers are more likely
to become AGN-dominated ULIRGs.

(4) For most models,  major mergers can become
ULIRGs, powered either by starbursts or by AGN, only when the two bulges
finally merge.  
Interacting disk galaxies can become
ULIRGs with well separated  two  cores ($>$ 20kpc) 
at their pericenter
when they are  very massive
and have  small bulges. 

(5) Irrespective of the choice of model,
interacting/merging galaxies show the highest accretion rates
onto the central SMBHs, and the resultant rapid growth of the SMBHs
occur when their star formation rates are very high.
 
Based on these results, we discuss an evolutionary link between ULIRGs,
QSOs with poststarburst populations, and ``E+A'' galaxies. We also
discuss spectroscopic properties (e.g., H$\beta$ luminosities and
line ratio of \oiii/H$\beta$) in galaxy mergers with starbursts and
AGN.

\end{abstract}


\section{Introduction}

The observation that ultra-luminous infrared galaxies (ULIRGs,
defined as those with infrared luminosities
greater than  $10^{12}$ $ L_{\odot}$) show a mixture of two distinct types of
nuclear activities, namely starbursts and
active galactic nuclei (AGN), has led to many observational studies
of their formation and evolution processes
(e.g., Sanders et al. 1988; Solomon et al. 1992; 
Soifer et al.  1986; Clements et al. 1996;
Murphy et al. 1996; Sanders \& Mirabel 1996; 
Gao \& Solomon 1999; Trentham et al. 1999;
Veilleux et al. 1999; Scoville et al. 2000; Surace et al. 2000;
Bushouse et al. 2002; Tacconi et al. 2002;
Farrah et al. 2003; 
Armus et al. 2004; Imanishi \& Terashima 2004; 
Colina et al. 2005; Iwasawa et al. 2005).
For example, 
Sanders et al. (1988)  proposed that  ULIRGs formed by gas-rich
galaxy mergers can finally evolve into QSOs after the  removal
of dust surrounding  QSO black holes.
Spectroscopic properties of ULIRGs have been extensively discussed
in terms of   
the relative importance of starbursts and 
active galactic nuclei (AGN) in the energy budget of
ultra-luminous infrared galaxies
(e.g., Genzel et al. 1998; Lutz et al. 1998).

These observations have so far raised many questions,
the most significant being:
(1) whether all ULIRGs   evolve into QSOs,
(2) what mechanisms are responsible for triggering 
starbursts  and AGN obscured heavily by dust in ULIRGs,
(3) what determines the relative
importance of 
starbursts and AGN  in spectral energy distributions (SEDs) of
ULIRGs, 
(4) whether there is an evolutionary link between
starbursts and AGN in ULIRGs, 
and (5) whether there can be physical relationships
between low redshift (low-$z$) ULIRGs and high-$z$ 
dust-enshrouded starbursts
and AGN at intermediate and high redshifts 
recently revealed by SCUBA (Submillimeter Common-User Bolometer Array)
(e.g., Barger et al. 1998;
Smail et al. 1997, 1998, 1999; Blain et al. 1999). 
Morphological studies of ULIRGs revealed that 
they show strongly disturbed morphologies indicative
of violent galaxy interaction and merging.
Previous theoretical studies have tried to answer  
the above five questions
in the context of gas fuelling to the central
region of galaxy mergers (See Shlosman et al. 1990
for more general discussions on fuelling mechanism in
galaxies).

Physical mechanisms responsible for the formation of 
starbursts in galaxy mergers have been investigated
by many authors  
(e.g., Olson \& Kwan 1990; Barnes \& Hernquist 1991;
Noguchi 1991;
Mihos \& Hernquist 1994, 1996; 
Gerritsen \& Icke 1997).
For example, Olson \& Kwan (1990) suggested that
high velocity disruptive cloud-cloud collisions,
which are more prominently enhanced in mergers,
are responsible for the observed high star formation 
rates in galaxy mergers. 
Although these previous numerical studies provided some
theoretical predictions on star formation rates 
(SFRs) and their dependence
on the initial physical parameters of galaxy merging 
(e.g., bulge-to-disk-ratio and gas mass fraction),
they did not investigate both SFRs and accretion rates (ARs) 
onto the central super-massive black holes (SMBHs)  simultaneously. 
Therefore, they did not provide  useful theoretical predictions
on the formation and evolution of AGN, or
on a possible evolutionary link between starbursts and AGN
in ULIRGs.

Physical processes of  gas fuelling to 
the central SMBHs in galaxy mergers
have been investigated by a number of authors 
(Bekki \& Noguchi 1994; Bekki 1995; Di Matteo et al. 2005;
Springel et al. 2005a, b).
Using dynamical simulations with rather idealized modeling
of gas dynamics and star formation, 
Bekki \& Noguchi (1994) first investigated both 
%star formation rates (SFRs) and accretion rates onto SMBHs (ARs)
SFRs and ARs
in merging galaxies 
and found that 
SFRs become very high at the epoch of
the coalescence of the cores of two merging galaxies,
whereas ARs attain their maxima only after the coalescence.  
Recently, Springel et al (2005a) have performed more sophisticated,
high-resolution
SPH simulations including feedback effects of AGN on
the interstellar medium (ISM), and thereby demonstrated that AGN
feedback can be quite important
for global photometric properties of elliptical galaxies 
formed by major galaxy merging.  
These previous models however did not discuss
the latest observational results of ULIRGs,
partly because their model do not allow authors to investigate
photometric and spectroscopic properties of dusty starbursts
and AGNs in galaxy mergers.

The purpose of this paper is thus to investigate simultaneously
both SFRs and ARs
of merging galaxies in an self-consistent manner
and thereby try to address  the aforementioned
questions related to the origin of ULIRGs.
We particularly try to understand (1) physical conditions
required for galaxy mergers to evolve into ULIRG with
AGN (or starbursts), (2) key factors which determine the relative
importance of starbursts and AGN,
and (3) epochs when mergers become ULIRGs with AGN. 
We  develop a new model in which the physics of star formation
(including gas consumption and supernovae feedback by star formation),
the time evolution  of accretion
disks around SMBHs,  and the growth of SMBHs via gas accretion
from the accretion disks are included.
By using this new model,
we  show (1) how SFRs and ARS in merging galaxies
evolve with time,
(2) how they depend on 
galactic masses, mass ratios of two merging spirals,
and bulge-to-disk-ratios of the merger progenitor spirals,
and (3) how SMBHs grow
in the central regions of starbursting mergers.
We also show emission  line properties
of galaxies with starbursts and AGNs  by combining the results of
the simulated SFRs and ARs
with spectral evolution codes.

Although previous numerical simulations combined with
spectrophotometric synthesis codes have already
derived SEDs of {\it purely starburst} galaxies obscured by dust
(Bekki et al. 1999; Bekki \& Shioya 2000, 20001; Jonsson et al. 2005),
they did not discuss at all the spectrophotometric properties
of galaxies {\it where starbursts and AGN coexist}.
Therefore our new way of deriving  spectral properties
based on simulation results  enables us to answer some key questions 
raised by recent large, systematic survey of AGN
(e.g., Kauffmann et al. 2003),
such as
why a significant fraction
of high-luminosity AGN have the  Balmer absorption lines.
Previous one-zone spectroscopic models discussed
what controls emission and absorption line properties
of galaxies with starbursts and AGN 
(Baldwin, Phillips \& Terlevich 1981; Veilleux \& Osterbrock 1987;
Kewley et al. 2001; Dopita et al. 2006).
The present simulations allow us to discuss this point
based on the results of SFRs and ARs derived by chemodynamical
simulations with growth of SMBHs.

The plan of the paper is as follows: In the next section,
we describe our numerical model for calculating SFRs and
ARs in merging galaxies. 
In \S 3, we
present the numerical results
on the time evolution of SFRs and ARs and its dependences
of model parameters.
In this section, we also show emission line properties of galaxies mergers 
with starbursts and AGN.
We discuss the present results in  terms of formation and evolution
of ULIRGs and QSOs 
in \S 4.
We summarise our  conclusions in \S 5.

%%%%%%TABLE

\begin{table*}
\centering
\begin{minipage}{185mm}
\caption{Model parameters}
\begin{tabular}{ccccccccl}
{Model no. } & 
$M_{d}$ ($\times$ $10^{10} {\rm M_{\odot}} $)   & 
{$f_{\rm g}$% 
\footnote{initial gas mass fraction}} &
{$f_{\rm b}$%
\footnote{mass ratio of bulge to disk}} &
{$m_{\rm 2}$%
\footnote{mass ratio of merging two disks}} &
{orbital type} & 
{$m_{\rm sf,max}$%
\footnote{maximum star formation  rate (${\rm M}_{\odot}$ yr$^{-1}$)}} &
{$m_{\rm acc,max}$%
\footnote{maximum accretion rate (${\rm M}_{\odot}$ yr$^{-1}$)}} &
Comments \\
M1 & 6.0 & 0.2 & 0.5 & 1.0 & FI & $2.6 \times 10^0$  & $2.5\times 10^{0}$ & standard  \\
M2 & 6.0 & 0.2 & 0.5 & 1.0 & FI & $6.4 \times 10^2$& $0\times 10^{0}$ & no accretion onto SMBHs  \\
M3 & 0.15 & 0.2 & 0.5 & 1.0 & FI & $1.3 \times 10^{-1}$ & $4.0 \times 10^{-4}$ &  \\
M4 & 3.0 & 0.2 & 0.5 & 1.0 & FI & $5.0\times 10^0$ & $7.0 \times 10^{-1}$ &  \\
M5 & 30.0 & 0.2 & 0.5 & 1.0 & FI & $3.0 \times 10^2$ & $9.2\times 10^1$ &  \\
M6 & 0.15 & 0.2 & 0.1 & 1.0 & FI & $2.1 \times 10^{-1}$ & $6.9 \times 10^{-5}$  &  \\
M7 & 3.0 & 0.2 & 0.1 & 1.0 & FI & $5.1\times 10^0$ & $8.2 \times 10^{-3}$ &  \\
M8 & 6.0 & 0.2 & 0.1 & 1.0 & FI & $9.5\times 10^0$ & $3.0 \times 10^{-2}$ &  \\
M9 & 30.0 & 0.2 & 0.1 & 1.0 & FI & $9.8 \times 10^1$ & $9.5 \times 10^{-1}$ &  \\
M10 & 6.0 & 0.2 & 0.5 & 1.0 & HI & $2.2 \times 10^1$ & $2.3\times 10^0$ &  \\
M11 & 6.0 & 0.2 & 0.5 & 1.0 & RR & $2.2 \times 10^1$ & $3.6\times 10^0$ &  \\
M12 & 6.0 & 0.2 & 0.5 & 1.0 & BO & $3.3 \times 10^1$ & $4.1\times 10^{0}$ &  \\
M13 & 6.0 & 0.2 & 0.5 & 0.1 & BO & $6.0 \times 10^0$ & $1.0\times 10^{-2}$ & LSB minor merger  \\
M14 & 6.0 & 0.2 & 0.5 & 0.3 & BO & $6.8 \times 10^0$ & $1.0\times 10^{-1}$ & unequal-mass merger  \\
M15 & 6.0 & 0.2 & 0.5 & 0.1 & BO & $8.5 \times 10^0$  & $2.0\times 10^{-1}$ & HSB minor  merger  \\
M16 & 6.0 & 0.02 & 0.5 & 1.0 & FI & $6.8 \times 10^{-1}$ & $4.9 \times 10^{-3}$ &  gas poor \\
M17 & 6.0 & 0.05 & 0.5 & 1.0 & FI & $1.8 \times 10^0$ & $1.0 \times 10^{-1}$ &   \\
M18 & 6.0 & 0.1 & 0.5 & 1.0 & FI & $8.3 \times 10^0$ & $1.6 \times 10^0$ &   \\
M19 & 6.0 & 0.2 & 0.02 & 1.0 & FI & $1.1 \times 10^2$ & $2.9 \times 10^{-3}$ & smaller bulge \\
M20 & 6.0 & 0.2 & 1.0 & 1.0 & FI & $6.0 \times 10^1$ & $1.6 \times 10^0$ &  bigger bulge\\
M21 & 6.0 & 0.05 & 1.0 & 1.0 & FI & $3.1 \times 10^0$ & $6.1 \times 10^{-1}$ & bigger bulge, gas poor \\
M22 & 30.0 & 0.2 & 0.1 & 1.0 & TI & $5.7 \times 10^1$ & $2.4 \times 10^0$ & tidal interaction  \\

\end{tabular}
\end{minipage}
\end{table*}

\begin{figure*}
\psfig{file=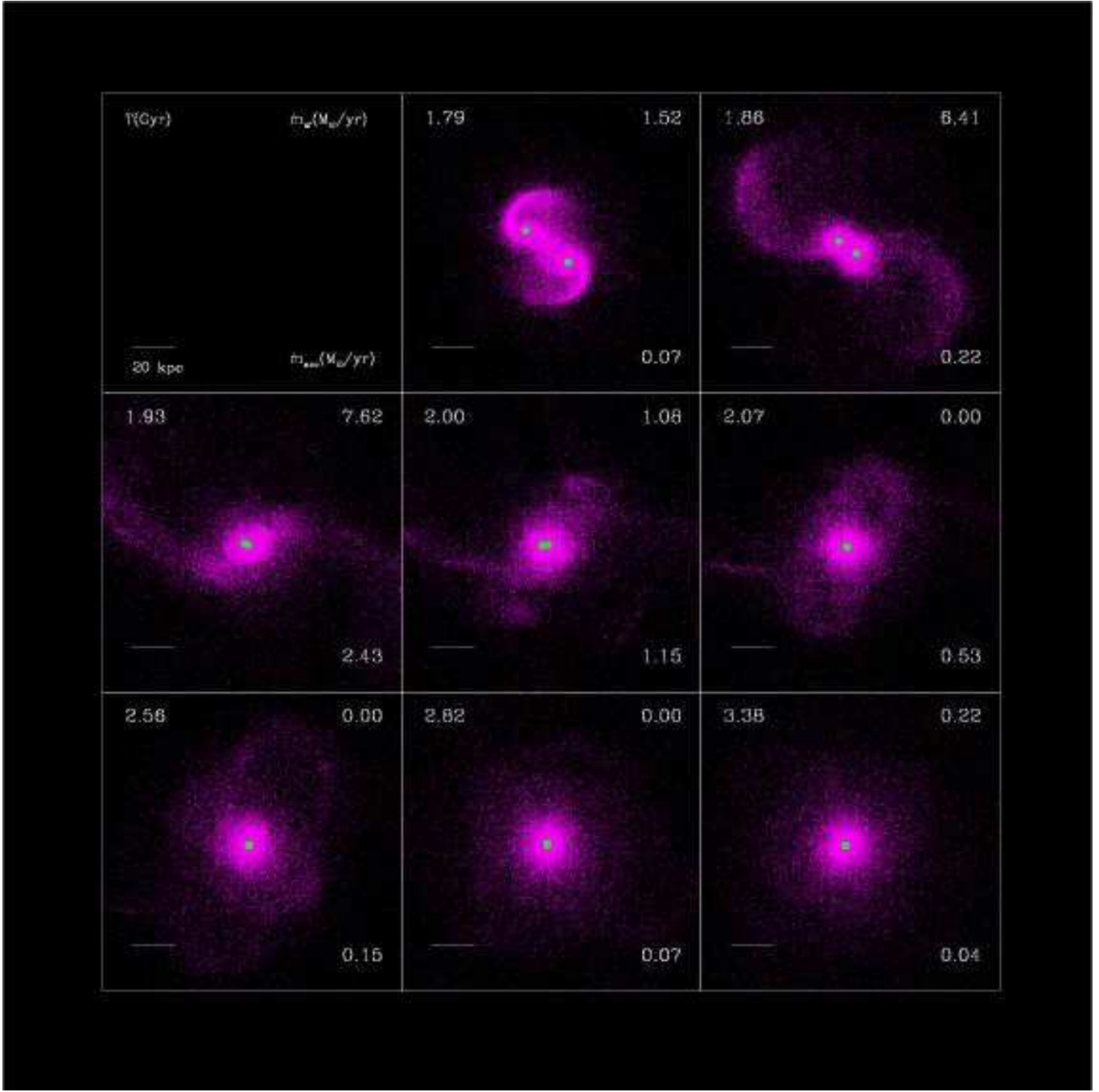,width=18.0cm}
\caption{ 
Mass distributions projected onto the $x$-$y$ plane
for the standard model. For convenience,   
stellar particles (old stars) and gaseous ones are shown
in magenta (i.e.\ dark matter halo particles are not shown).
Big green dots represent the locations of SMBH1 and 2.
Time ($T$), 
SFRs (${\dot{m}}_{\rm sf}$ in units of ${\rm M}_{\odot} {\rm yr}^{-1}$), 
ARs (${\dot{m}}_{\rm ac}$ in units of ${\rm M}_{\odot} {\rm yr}^{-1}$) 
and the simulation scale
are shown at upper left, upper right, lower right, and lower left,
respectively, for each frame.
Here time $T$ represents 
the time that has elapsed since
the simulation starts.
Note that ARs can become very high ($> 1{\rm M}_{\odot} {\rm yr}^{-1}$)
when the two bulges finally merge (i.e.\ when the two SMBHs become
very close with each other).
}
\label{Figure. 1}
\end{figure*}

\section{The model}

Since the numerical methods and techniques we employ for modeling the chemodynamical
and photometric evolution of galaxy mergers have already been described in detail 
elsewhere (Bekki \& Shioya 1998, 1999), we give only  a brief review here. 

\subsection{Progenitor disk galaxies}

The progenitor disk galaxies that take part in a merger are taken to 
have a dark halo, a bulge, and a thin exponential disk.
Their total mass and size are $M_{\rm d}$ and $R_{\rm d}$, respectively. 
From now on, all masses are measured in units of
$M_{\rm d}$ and  distances in units of $R_{\rm d}$, unless otherwise specified. 
Velocity and time are measured in units of $v$ = $ (GM_{\rm d}/R_{\rm d})^{1/2}$ and
$t_{\rm dyn}$ = $(R_{\rm d}^{3}/GM_{\rm d})^{1/2}$, respectively,
where $G$ is the gravitational constant and assumed to be 1.0
in the present study. 
If we adopt $M_{\rm d}$ = 6.0 $\times$ $10^{10}$ $ \rm M_{\odot}$ and
$R_{\rm d}$ = 17.5\,kpc as fiducial values, then $v$ = 1.21 $\times$
$10^{2}$\,km\,s$^{-1}$  and  $t_{\rm dyn}$ = 1.41 $\times$ $10^{8}$ yr.

We adopt the density distribution of the NFW 
halo (Navarro, Frenk \& White 1996) suggested from CDM simulations:
 \begin{equation}
 {\rho}(r)=\frac{\rho_{s}}{(r/r_{\rm s})(1+r/r_{\rm s})^2},
 \end{equation} 
 where  $r$, $\rho_{s}$, and $r_{\rm s}$ are
the spherical radius,  the characteristic
  density of a dark halo,  and the scale
length of the halo, respectively.  
The dark matter distribution is truncated at
$r=10r_{\rm s}$ corresponding to $r_{200}$ in the NFW.
The value of $r_{\rm s}$ (0.8) is chosen such that
the rotation curve of the disk is reasonably consistent with
observations. The bulge has a density profile
with a shallow cusp (Hernquist 1990): 
 \begin{equation}
 {\rho}(r) \propto r^{-1}(r+a_{\rm bulge})^{-3},
 \end{equation} 
where $a_{\rm bulge}$ is the scale length of the bulge.
The ratio of a bulge mass ($M_{\rm b}$) to a disk mass ($M_{\rm d}$)
in a disk is regarded as a free parameter and represented as $f_{\rm b}$.
We determine 
the bulge scale length, $a_{\rm bulge}$, for a given $M_{\rm b}$
based on the Faber-Jackson relation (Faber \& Jackson 1976) and
the virial theorem.
The bulge mass  and its compactness can control the bar formation
in the disks and thus the strength of starbursts in mergers. 
The bulge contains a SMBH with the mass ($M_{\rm SMBH}$) following the observed
relation (Magorrian et al. 1998);
\begin{equation}
M_{\rm SMBH}=0.006 M_{\rm b}=0.006f_{\rm b}M_{\rm d}.
\end{equation}
The model for the time evolution of $M_{\rm SMBH}$ is described later.

The dark matter to disk mass ratio 
is fixed at 9 whereas the bulge to disk ratio 
is assumed to be a free parameter ($f_{\rm b}$).
The radial ($R$) and vertical ($Z$) density profiles 
of the  disk are  assumed to be
proportional to $\exp (-R/R_{0}) $ with scale length $R_{0}$ = 0.2
and to  ${\rm sech}^2 (Z/Z_{0})$ with scale length $Z_{0}$ = 0.04
in our units, respectively.
In addition to the rotational velocity attributable to the gravitational
field of the disk and halo components, the initial radial and azimuthal velocity
dispersions are added to the disk component in accordance with
the epicyclic theory, and with a Toomre parameter value of $Q$ = 1.5
(Binney \& Tremaine 1987) .
The vertical velocity dispersion at a given radius 
is set to be 0.5 times as large as
the radial velocity dispersion at that point, 
as is consistent with the trend observed in the Milky Way (e.g., Wielen 1977).

The ratio of  $R_{0}$ to $R_{d}$ and that of $r_{\rm s}$ to $R_{0}$
are fixed at 0.2 and 4.0, respectively, for all disk models with
different $M_{\rm d}$. Since we adopt the scaling relation of
$ {\mu}_{s} \propto {M_{\rm d}}^{0.5} $ (Kauffmann et al. 2003b),
where $ {\mu}_{s} $ is  the mean stellar surface density of a disk
(described later in 2.6),  $ R_{\rm d} = C_{\rm s} 
\times {M_{\rm d}}^{0.25} $ 
(or  $ R_{\rm 0} \propto  {M_{\rm d}}^{0.25} $). 
The normalization factor $C_{\rm s}$ is determined such that
$ R_{\rm 0}=3.5$ kpc for 
$M_{\rm d}$ = 6.0 $\times$ $10^{10}$ $ \rm M_{\odot}$.
Thus the scale lengths
of disks are different between models with different $M_{\rm d}$.

\subsection{Star formation rates}

The disk is composed both of gas and stars, with the gas mass fraction
($f_{\rm g}$) being a free parameter and the gas disk
represented by a collection of discrete gas clouds that follow the observed mass-size
relationship (Larson 1981). All overlapping pairs of gas clouds
are made to collide with the same restitution coefficient of 0.5
(Hausman \& Roberts 1984). The gas is converted solely into field stars:
we do not consider the formation of globular clusters (GCs).
Field star formation
is modeled by converting  the collisional gas particles
into  collisionless new stellar particles according to the algorithm
of star formation  described below. We adopt the Schmidt law (Schmidt 1959)
with exponent $\gamma$ = 1.5 (1.0  $ < $  $\gamma$
$ < $ 2.0, Kennicutt 1998) as the controlling
parameter of the rate of star formation. The amount of gas 
consumed by star formation for each gas particle
in each time step is given by:
\begin{equation}
\dot{{\rho}_{\rm g}} \propto  
{\rho_{\rm g}}^{\gamma},
\end{equation}
where $\rho_{\rm g}$ 
is the gas density around each gas particle. 
The coefficients in the law are taken from the work of Bekki (1998, 1999): 
The mean star formation rate in an isolated disk model
with $M_{\rm d}$ = 6.0 $\times$ $10^{10}$ $ \rm M_{\odot}$ 
and the gas mass fraction of 0.1 
for 1 Gyr evolution is $\sim$ 1 ${\rm M}_{\odot}$ for the adopted
coefficient (thus consistent with the observed star formation 
rate in the Galaxy; e.g., van den Bergh 2000).
These field stars formed from gas are called ``new stars'' (or ``young stars'')
whereas stars initially within a disk are called ``old stars'' 
throughout this paper.
The adopted star formation model is similar to that with
$C_{\rm SF}=3.5$ in Bekki \& Shioya (1998).

Chemical enrichment through star formation during galaxy merging
is assumed to proceed both locally and instantaneously in the present study.
We assign the metallicity of the original
gas particles to  the new stellar particles and increase 
the metallicity of the neighboring gas particles. 
The total number of neighboring gas particles is taken to be $N_{\rm gas}$,
given by the following equation for chemical enrichment:
  \begin{equation}
  \Delta M_{\rm Z} = \{ Z_{i}R_{\rm met}m_{\rm s}+(1.0-R_{\rm met})
 (1.0-Z_{i})m_{\rm s}y_{\rm met} \}/N_{\rm gas}. 
  \end{equation}
Here, $\Delta M_{\rm Z}$ represents the increase in metallicity for each
gas particle, $ Z_{i}$ the metallicity of the new stellar particle (or that
of the original gas particle), $R_{\rm met}$ the fraction of gas returned
to the interstellar medium, $m_{\rm s}$ the mass of the new star,
and $y_{\rm met}$ the chemical yield.
The values of $R_{\rm met}$ and $y_{\rm met}$ are set to 0.3 and 0.02 respectively.

Using numerical simulations, Thornton et al. (1998) demonstrated that 
the total amount of energy that supernovae can give to the ISM ranges from
$\approx 9 \times 10^{49}$ to $\approx 3 \times 10^{50}$ ergs with
a typical case being $\approx  10^{50}$ ergs. 
This amount  is roughly 10 \% of the total amount of
energy of Type II SN and 20 \% of Type I (Thornton et al. 1998).
They also found that most of the energy of supernovae can be in the form of 
kinetic energy within the ISM. Guided by these previous theoretical results,
we assume that 10\% of supernovae energy
can be converted into kinematical energy
of gas clouds. We adopt the Salpeter IMF with the lower mass cut off of
$0.1\,{\rm M}_{\odot}$, the upper one of $100.0\,{\rm M}_{\odot}$
and the exponent of the slope equal to $-2.35$ (i.e.\ a canonical IMF).
Total number of supernovae at each time step can be calculated according
to the star formation rate. The more details of the numerical method to
give kinematical energy of supernovae to gas clouds 
 are given in Bekki \& Shioya (1999).

\subsection{Accretion rates onto SMBHs}

AGN activity is believed to originate from sub-parsec
size regions at the galactic nuclei, powered by the mass accretion onto
SMBHs through accretion disks (e.g., Rees 1984). 
In the present study, we assume that SMBHs are not surrounded
by accretion disks in initial disks and thus the accretion disks
are assumed to form during merging.
Therefore, we need to model (1) formation processes of accretion
disks around SMBHs and (2) time evolution of ARs in
growing accretion disks 
in order to estimate ARs in an self-consistent manner.
Although numerous theoretical studies have already been made for physical properties
of static accretion disks (e.g., Frank, King \& Raine 2002),
there have been  no extensive theoretical studies on ARs in
accretion disks that are {\it forming and growing} through radial gas inflow
into the central sub-parsec-scale region of galaxies from their outer parts. 
Furthermore only a few theoretical attempts have been made to elucidate
the formation process of gaseous tori and accretion disks around SMBHs 
(e.g., Bekki 2000).

%Considering this situation of theoretical studies,
Given this lack of theoretical detail on the evolution of accretion disks,
we adopt the
following two-fold model to calculate ARs.
For each time step of a simulation,
we first calculate total gas mass that can be used for the formation 
of an accretion disk around a SMBH in the central region of a galaxy 
by assuming that tidal interaction between the SMBH and its nearby gas clouds
and the resultant destruction of gas clouds (Bekki 2000) can be 
responsible for gas supply to the accretion disk.
Then, by using a reasonable analytical model, 
we calculate the time evolution of the AR ($\dot{m}_{\rm acc}$) 
onto the SMBH for a given mass of the accretion disk at each time step.
The details of this two-fold model are described as follows.

\subsubsection{Formation of an accretion disk around a SMBH}

The mass of an accretion disk around a SMBH is assumed to
increase as a result of gas accretion from gas clouds being
gravitationally trapped and destroyed by the SMBH (Bekki 2000).
We estimate the total mass of an accretion disk ($M_{\rm ad}$) around
a SMBH based on gas densities of gas clouds within $R_{\rm acc}$ 
from the SMBH. We assume that gas clouds within $R_{\rm acc}$ can be 
used as fuel for an accretion disk. Bekki \& Noguchi (1994) adopted
$R_{\rm G}$ as  $R_{\rm acc}$, where $R_{\rm G}$ is defined as
\begin{equation}
R_{\rm G}=\frac{GM_{\rm SMBH}}{{\sigma}^2}
\end{equation}
where $M_{\rm SMBH}$ is the mass of the SMBH and
${\sigma}$ is the velocity dispersion (or any characteristic
velocity) in the background components, and G is the gravitational
constant. This $R_{\rm G}$  can be estimated to be $\sim 10$pc 
for $M_{\rm SMBH}=10^8 {\rm M}_{\odot}$
in a canonical set of galaxy parameters (Bekki \& Noguchi 1994).

Since our model for $M_{\rm ad}$ evolution is based on
interaction between SMBHs and gas clouds,
we adopt the ``Bondi'' radius ($R_{\rm B}$) rather than $R_{\rm G}$ 
as $R_{\rm acc}$. 
$R_{\rm B}$ is described as
\begin{equation}
R_{\rm B}=\frac{2GM_{\rm SMBH}}{{v_{\rm rel}}^2}
\end{equation}
where $v_{\rm rel}$
is relative velocity between the SMBH and gas.
We assume that $v_{\rm rel}$ is equivalent to the central velocity dispersion
of a bulge in each model. Therefore, $R_{\rm acc}$
is initially determined by the bulge mass of $M_{\rm b}$ ($=f_{\rm b}M_{\rm d}$)
in a model owing to the adopted relation 
of $M_{\rm SMBH}=0.006M_{\rm b}$ relation.

Next we estimate total mass of gas clouds within $R_{\rm acc}$ and
thereby derive a local gas density (${\rho}_{\rm g}$) around a SMBH. 
Guided by theoretical predictions
by Hoyle \& Lyttleton (1941),  Bondi (1952), and Ruffert \& Arnett (1994),
we calculate the accretion disk from gas clouds (${\dot{m}}_{\rm cl}$)
as follows:
\begin{equation}
{\dot{m}}_{\rm cl}=C_{\rm B} \times \rho_{\rm g},
\end{equation}
where $C_{\rm B}$ is linearly proportional to ${M_{\rm SMBH}}^{2}$ and
${v_{\rm rel}}^{-3}$.
$C_{\rm B}$ is therefore chosen according to $M_{\rm SMBH}$ and
$M_{\rm b}$ (and hence $M_{\rm d}$  and  $f_{\rm b}$) in each model.
Time evolution of $M_{\rm ad}$ is determined by 
solving the following equation in terms of ${\dot{m}}_{\rm cl}$
and ${\dot{m}}_{\rm acc}$:
\begin{equation}
{\dot{M}}_{\rm ad}={\dot{m}}_{\rm cl}-{\dot{m}}_{\rm acc}.
\end{equation}
${\dot{m}}_{\rm acc}$ represents the mass accretion rate onto
a SMBH and 
we describe the way to estimate ${\dot{m}}_{\rm acc}$ later.
Equation (9) thus means that if there is no supply of gas from
outer part of a galaxy into the nuclear accretion disk,
$M_{\rm ad}$ gradually decreases owing to consumption of
gas within the disk.

\subsubsection{Time evolution of accretion rates}

Based on $M_{\rm ad}$, we estimate AR (${\dot{m}}_{\rm acc}$) at each
time step in a simulation. 
We adopt a gas-pressure dominated standard $\alpha$-disk with
the conversion rate of accreted mass to energy($\epsilon$) equal to 0.1
and follow the
relation between $M_{\rm ad}$ and ${\dot{m}}_{\rm acc}$ for a
given $M_{\rm SMBH}$ shown in Liu (2004):
\begin{equation}
{\dot{m}}_{\rm acc}=0.1{(\frac{M_{\rm ad}}{7.7\times 10^7{\rm M}_{\odot}})}^{5/3} 
{\rm M}_{\odot} {\rm yr}^{-1}.
\end{equation}
We assume that  ${\dot{m}}_{\rm acc}$ should not exceed the Eddington accretion rate,
which is described as:
\begin{equation}
{\dot{m}}_{\rm Edd}=2.3(\frac{M_{\rm SMBH}}{10^8{\rm M}_{\odot}})
{\rm M}_{\odot} {\rm yr}^{-1}.
\end{equation}
Thus, if ${\dot{m}}_{\rm acc} > {\dot{m}}_{\rm Edd}$, ${\dot{m}}_{\rm acc}$
is set to be ${\dot{m}}_{\rm Edd}$ at every time step in all simulations.

As a result of gas accretion onto a SMBH, $M_{\rm SMBH}$ is time-dependent
and its evolution is described as:  
\begin{equation}
M_{\rm SMBH}(t+\Delta t)=M_{\rm SMBH}(t)+{\dot{m}}_{\rm acc}\times \Delta t,
\end{equation}
where $\Delta t$ is the time step width (corresponding to $0.01t_{\rm dyn}$)
in a simulation.
We consider that 10\% of the rest mass of accreted gas 
(i.e.\ $0.1{\dot{m}}_{\rm acc}\Delta t$) can be converted into energy ($E_{\rm acc}$).
Although some fraction of $E_{\rm acc}$ may well be used for thermally and
dynamically heating the ISM in galaxy mergers,
it is unclear what fraction ($f_{\rm agn}$) of $E_{\rm acc}$ can be returned back
to the ISM through AGN feedback effects.
Accordingly, we compromise by assuming that $f_{\rm agn}=0.1$,
the same value as that derived 
for supernovae feedback effects (Thornton et al. 1998).

The AGN feedback energy from  a SMBH  is assumed to be 
used for the increase of kinetic energy of gas particles
around the SMBH. Therefore, $f_{\rm agn} E_{\rm acc}$ 
is equivalent to the sum of the increase in  kinematical 
energy of gas particles at each time step. The methods
to give a velocity perturbation (directed radially 
away from the SMBH) to each gas particle around the SMBH
are the same as those for stellar feedback effects in
Bekki \& Shioya (1999). The present results depend
on $f_{\rm agn}$ such that gas transfer to nuclei
(thus nuclear star formation and AGN fueling) can be
more strongly suppressed in the models with larger $f_{\rm agn}$.
In this paper, we consider that the adopted value of 0.1
is reasonable, because this value is similar both
to that in Springel et al. (2005a) with $f_{\rm agn}=0.05$
(explaining  some observations) and to that by
Thornton et al. (1998) for kinetic feedback effects
for supernovae.

\begin{figure}
\psfig{file=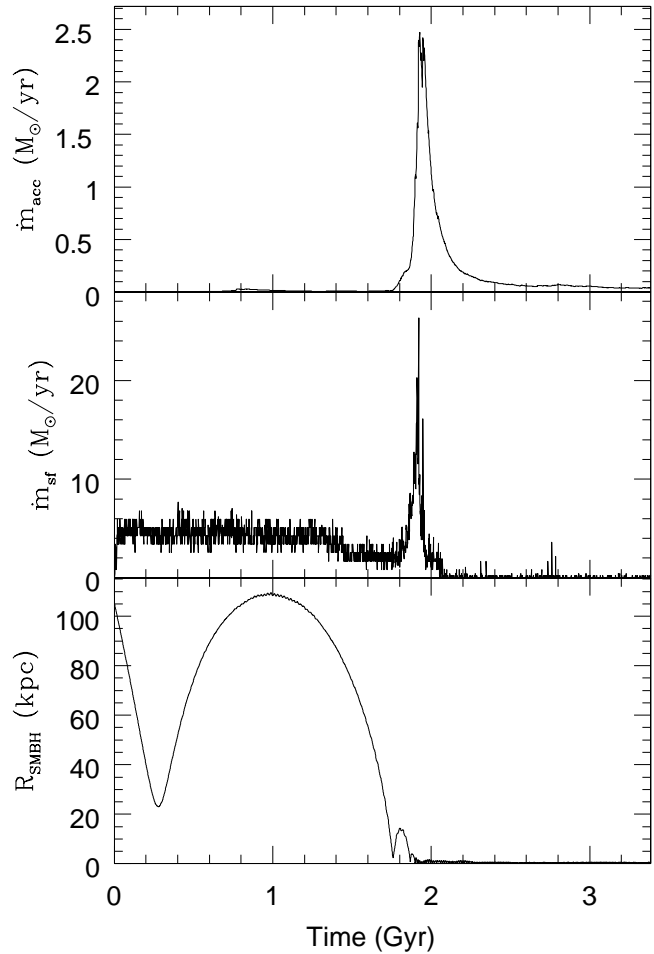,width=8.5cm}
\caption{ 
Time evolution of  AR (${\dot{m}}_{\rm acc}$; top),
SFR (${\dot{m}}_{\rm sf}$; middle),
and separation of two SMBHs ($R_{\rm SMBH}$; bottom)
in the standard model (M1).
Note that the peak of the AR is nearly coincident with 
that of the SFR.
}
\label{Figure. 2}
\end{figure}

\subsection{Orbital configurations}

In all of the simulations of merging pairs, the orbit of the two disks is set to be
initially in the $xy$ plane and the distance between
the center of mass of the two disks ($r_{\rm p}$)
is  assumed  to be ten times the disk size. 
The pericenter distance and the eccentricity
are set to be the disk size  and 1.0 (i.e.\ parabolic), 
respectively, for most of the models.
The spin of each galaxy in a merger
is specified by two angles $\theta_{i}$ and
$\phi_{i}$, where suffix  $i$ is used to identify each galaxy.
$\theta_{i}$ is the angle between the $z$ axis and the vector of
the angular momentum of a disk.
$\phi_{i}$ is the azimuthal angle measured from the $x$ axis to
the projection of the angular momentum vector of a disk onto the $xy$ plane.

We specifically present the results of the following three 
parabolic models with different disk inclinations with respect to the
orbital plane: A fiducial  model represented by ``FI''
with $\theta_{1}$ = 0, $\theta_{2}$ = 30, $\phi_{1}$ = 0, 
and $\phi_{2}$ = 0;
a retrograde-retrograde model (``RR'') with $\theta_{1}$ = 180,
$\theta_{2}$ = 210, $\phi_{1}$ = 0, and $\phi_{2}$ = 0; and
a highly inclined model (``HI'') with $\theta_{1}$ = 60, $\theta_{2}$ =
60, $\phi_{1}$ = 90, and $\phi_{2}$ = 0.
In addition to these parabolic models with $e_{\rm p}=1$, 
the bound orbit model (``BO'') with the orbital eccentricity of 0.5
and with the same orbital configuration
as the ``FI'' model is investigated.
The time taken for the progenitor disks to completely merge and reach 
dynamical equilibrium is less than 16.0 in our units ($\sim$ 2.2\,Gyr) for most of
our major merger models. 

In order to compare the time evolution of SFRs and ARs in mergers with
that of tidally interacting galaxies,
we investigate tidal interaction models (``TI'').
Although we derive the results for several interaction models, we show only the most 
interesting case in the present study, since our main
interest is on SFRs and ARs in galaxy mergers.
We show a  model with $e_{\rm p}=1.1$ (i.e.\ hyperbolic), 
$\theta_{1}$ = 0, $\theta_{2}$ = 30, $\phi_{1}$ = 0, $\phi_{2}$ = 0
and $r_{\rm p}$ equal to 1.5 times the disk size.
In the tidal interaction model, two disks do not merge at all
and become separated from each other 
soon after their pericenter passage.

All the calculations related to the above chemodynamical evolution
have been carried out on the GRAPE board (Sugimoto et al. 1990)
at the Astronomical Data Analysis Center (ADAC)
at the National Astronomical Observatory of Japan. 
The gravitational softening parameter was fixed at 0.025 in our
units (0.44\,kpc). The time integration of
the equation of motion was performed by using the 2nd-order leap-frog method.
Since the masses of the bulge particle are set to be the same 
in all simulations,
the initial total particle number in each simulation depends
on the bulge mass.
The total particle numbers for dark matter halo, bulge, stellar disk,
and gaseous one in 
a   model with $f_{\rm b}$ = 1.0 
are 60000, 10000, 20000, 20000, respectively,
in the present study.

\subsection{Emission line properties}

Our previous chemodynamical models with spectrophotometric synthesis
codes for dusty starburst galaxies have already demonstrated that
major  mergers between gas-rich spirals can become ULIRGs with 
$L_{\rm ir} > 10^{12} {\rm L}_{\odot}$,
because the triggered 
nuclear starburst components can be very heavily obscured
by dust (Bekki et al. 1999; Bekki \& Shioya 2000, 2001; Bekki et al. 2001).
Since these previous studies have already described the details
of evolution from galaxy mergers into ULIRGs, we here do not intend
to discuss the formation processes of ULIRGs. 
We instead discuss optical emission properties of galaxy mergers
with both starbursts and AGN based on SFRs and ARs derived 
from chemodynamical simulations. In the present paper,
we discuss {\it global, averaged spectral properties} of
galaxy mergers
rather than the spatial difference of the properties.
Two-dimensional distributions of emission line properties in ULIRGs
will be discussed in our forthcoming papers 
(Bekki \& Shioya 2005, in preparation).

We mainly demonstrate the time evolution of
emission line properties  of H$\alpha$, H$\beta$,
\oiii, and [NII] 
of galaxy mergers  
by considering the energy contribution from both thermal (i.e.\ starburst)
and non-thermal (i.e.\ AGN) components.
From the time evolution of $\dot{m}_{\rm sf}$ of a merger,
we first derive the SED 
at each time step by using stellar population synthesis codes.
This first step is exactly the same as that adopted in our
previous one-zone chemo-photometric galaxy evolution models
(Shioya \& Bekki 1998; 2000; Shioya et al. 2001, 2002, 2004). 
Secondly, we derive the total luminosity of the H$\beta$ line 
($L_{\rm H\beta}$) from the production rate of ionizing photons
based on the derived SED. Thirdly, by adopting  a typical
value of the H$\alpha$-to-H$\beta$ ratio 
(i.e.\ ${\rm H}\alpha/{\rm H}\beta \sim 2.9$),
and the observed values of \oiii/H$\beta$
and  \oiii/H$\alpha$ in HII regions of nearby galaxies
(Kennicutt et al. 1989),
we derive $L_{\rm H\alpha}$, $L_{\rm [O\,III]}$, and $L_{\rm [N\,II]}$.

Fourthly, we calculate the bolometric luminosity 
($L_{\rm bol}$) of AGN in the merger
from the AR by assuming that the 
energy conversion efficiency ($\epsilon$) 
in the accretion disk around a SMBH is
0.1. Fifthly, by adopting a reasonable set of values
of   $L_{\rm [O\,III]}/L_{\rm bol}=1/300$,
$L_{\rm [O\,III]}/L_{\rm x}=0.01$, and $L_{\rm bol}/L_{\rm x}=30$
(Kraemer et al. 2004),
we calculate the $L_{\rm [O\,III]}$ value due to the AGN.
Sixthly, we derive $L_{\rm H\alpha}$, $L_{\rm [O\,III]}$, and $L_{\rm [N\,II]}$.
by adopting  typical
values  of  
${\rm H}\alpha/{\rm H}\beta$,
\oiii/H$\beta$,
and \oiii/H$\alpha$ in nearby galaxies with Seyfert spectra
(Kennicutt et al. 1989).
Finally, we calculate the  total luminosities of emission lines
by combining the luminosities from starburst and  AGN components. 
The effects of dust on spectroscopic properties of galaxy mergers
model are not accounted for in the present model.

\begin{figure}
\psfig{file=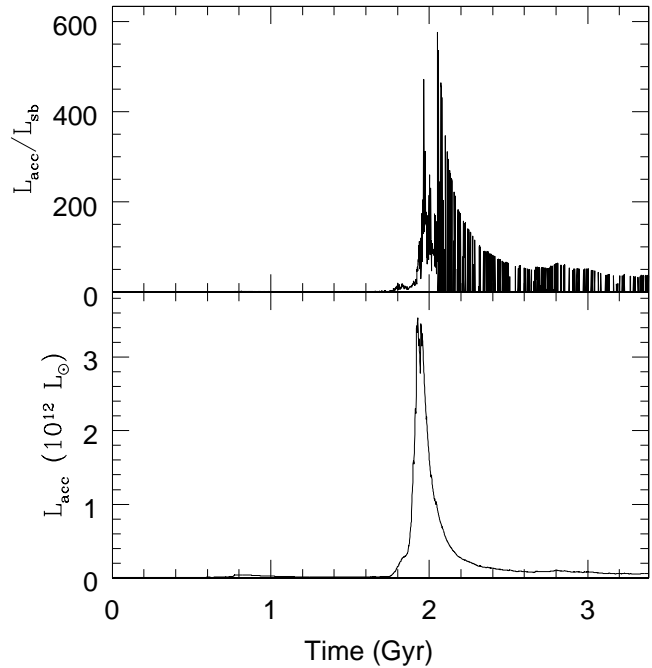,width=8.5cm}
\caption{ 
Time evolution of the ratio ($L_{\rm acc}/L_{\rm sf}$)
of bolometric luminosity coming from
AGN to that from starbursts (upper)  
and bolometric luminosity from AGN (lower)
in the standard model (M1).
Note that the time scale of the model to show the QSO-like
luminosity ($> 10^{12} {\rm L}_{\odot}$) is quite short ($\sim 0.1$Gyr).
Note also that $L_{\rm acc}/L_{\rm sf}$ becomes higher 
(i.e.\ galactic nuclei dominated by accretion power induced activities)
when $L_{\rm acc}$ becomes higher.  
}
\label{Figure. 3}
\end{figure}

\subsection{Main points of analysis}

We mainly investigate SFRs and ARs and their dependences on model
parameters of galaxy mergers. 
In order to clarify the relative importance of starbursts
and accretion power induced activities  in galactic nuclei, 
we estimate bolometric luminosities of starbursts ($L_{sb}$)
and AGN ($L_{\rm acc}$) by using analytic formula.
$L_{\rm sb}$ can be approximated as (Kennicut 1998):
\begin{equation}
L_{\rm sb}=10^{12}(\frac{{\dot{m}}_{\rm sf}}{140{\rm M}_{\odot} {\rm yr}^{-1}})
\times (\frac{{\eta}_{\rm nuc}}{0.01})
\times (\frac{f_{\rm ys}}{0.05})
{\rm L}_{\odot},
\end{equation}
where ${\eta}_{\rm nuc}$ is the
energy conversion efficiency in
nuclear fusion reactions (i.e.\ conversion of hydrogen to helium),
and $f_{\rm ys}$ is the fraction of massive young stars in starbursts. 
$L_{\rm acc}$ can be approximated as 
(Shapiro \& Teukolsky 1983; Frank et al 2002):
\begin{equation}
L_{\rm acc}=10^{12}(\frac{{\dot{m}}_{\rm acc}}{0.7{\rm M}_{\odot} {\rm yr}^{-1}})
\times (\frac{{\eta}_{\rm acc}}{0.1})
{\rm L}_{\odot},
\end{equation}
where ${\eta}_{\rm acc}$ is 
conversion efficiency of rest mass energy into radiation
in accretion disks. 
These equations (13) and (14) clearly mean that
for a galaxy to show a QSO-like luminosity  ($\approx 10^{12} {\rm L}_{\odot}$),
only small values of gas consumption rate ($0.7{\rm M}_{\odot} {\rm yr}^{-1}$)
are required if the QSO-like luminosity originates from accretion powered
activity. 

We investigate SFRs, ARs, and emission line properties of
galaxy mergers with starbursts and AGNs, and their dependences on
initial disk masses ($M_{\rm d}$), bulge-to-disk-ratios ($f_{\rm b}$),
gas mass fraction ($f_{\rm g}$), 
and mass ratios of two merging disks ($m_{\rm 2}$).
For the models with different $M_{\rm d}$ and those
with $m_{\rm 2} \neq 1$, we need to change masses and sizes
according to the scaling relation of galaxies. We adopt
the observed scaling relation by Kauffmann et al. (2003b)
and derive the following relation:
\begin{equation}
{\mu}_{s} \propto {M_{\rm d}}^{0.5},
\end{equation}
where ${\mu}_{s} $ is the mean stellar surface density of a disk.
We determine $R_{\rm d}$ for a given $M_{\rm d}$ 
by using the equation (15) and
the relation of ${\mu}_{s} \propto M_{\rm d}/{R_{\rm d}}^{2}$.
The above scaling relation means that less luminous galaxies
show lower surface brightness (LSB). 
For convenience,  
the model with $m_2$ = 0.1 (M13) 
including a smaller galaxy 
with the mass and the size consistent with
the equation 12 is referred to as a ``LSB  minor merger''.
We also investigate a ``HSB'' minor merger 
model with $m_2$ = 0.1 (M15) in which 
a smaller galaxy has a surface brightness 2 mag
higher than the LSB minor merger model.

We primarily show the results of the ``standard'' model M1, as this model
shows typical behavior for the evolution of SFRs and ARs.
Then we show the parameter dependences of other models. 
Below, we describe the results of 22 models and in Table 1 summarise
the model parameters for these: Model number (column 1),
total mass of a disk (2),
the gas mass fraction  (3),
the mass ratio of bulge to disk  (4),
the mass ratio $m_{2}$ of two merging disks  (5),
orbital types (6),
the maximum star formation rate  (7),
the maximum accretion rate  (8),
and comments on the models (9).
In the following discussion, the time $T$ represents the time that has elapsed since
the simulation starts.

\begin{figure}
\psfig{file=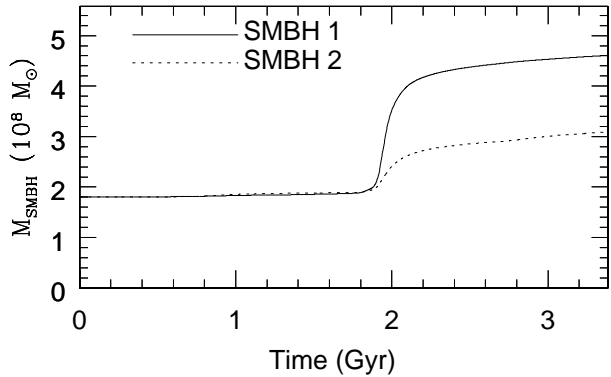,width=8.cm}
\caption{ 
The time evolution of SMBH1 (solid) and SMBH2 (dotted) in the standard model.
}
\label{Figure. 4}
\end{figure}

\begin{figure}
\psfig{file=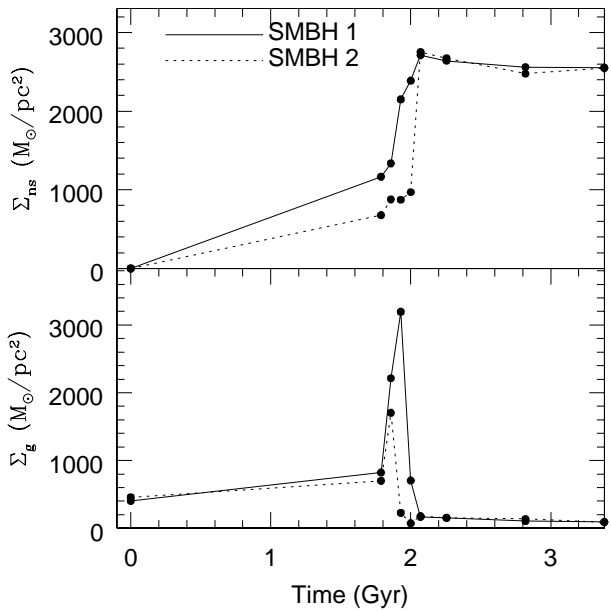,width=8.cm}
\caption{ 
Time evolution of projected mass density 
(${\Sigma}_{\rm ns}$) of new stars (i.e.\ poststarburst
populations) around SMBH1 (solid) and 2 (dotted) of the merger
(upper) and column gas density 
(${\Sigma}_{\rm g}$) around the SMBHs (lower) 
in the standard model. 
For clarity and comparison, only the results at nine epochs shown
in Figure 1 are described.
The projected stellar density and
column gas density are measured for particles that are located
within 0.04 in simulation units (corresponding to 0.7 kpc) around
SMBHs.
The significantly increased values of ${\Sigma}_{\rm ns}$
after $T \sim 1.8$ Gyr mean that SMBHs are surrounded by
compact poststarburst populations formed during galaxy merging.
The very large values of ${\Sigma}_{\rm g}$ around
$T=1.9$ Gyr mean
that SMBHs (thus AGN)  are heavily obscured  by metal-enriched
gas (thus dust).
}
\label{Figure. 5}
\end{figure}

\section{Results}

\subsection{The standard model}

\subsubsection{Evolution of SFRs and ARs}

Figure 1 illustrates the time evolution of morphological properties,
SFRs, and ARs simultaneously for the galaxy merger in the 
standard model.
The SFR can be moderately high 
(${\dot{m}}_{\rm sf} \sim 6.4 {\rm M}_{\odot} {\rm yr}^{-1}$)
at $T = 1.86$ Gyr when
two disks can be still clearly seen as separate entities,
whereas the AR becomes very high 
(${\dot{m}}_{\rm acc} > 1 {\rm M}_{\odot} {\rm yr}^{-1}$) 
at $T=1.93-2.00$ Gyr when two disks finally merge to form
a giant elliptical galaxy.
A disturbed outer morphology can be still seen 
at $T=2.07$ and 2.56 Gyr when the AR becomes lower
(i.e.\ weak AGN phases). Both the SFR and the AR become
significantly low at $T=2.82$ and 3.38 Gyr when
the merger remnant can be morphologically identified as an elliptical
with no peculiar fine structures (e.g., shells and plumes).

As Figure 2 reveals, showing the time evolution of SFR and AR in the standard model M1,
both SFR and AR can be maximised  when the two SMBHs become close to 
each other. This is essentially because efficient radial transfer of gas
into the central $10-100$pc in the merger occurs during the coalescence of
the two big bulges. The AR exceeds the $0.7 {\rm M}_{\odot} {\rm yr}^{-1}$ required
for QSO activity ($L_{\rm bol} > 10^{12} {\rm L}_{\odot}$) 
at $T= 1.9 \sim 2.0$ Gyr whereas the SFR does not exceed the
$100 {\rm M}_{\odot} {\rm yr}^{-1}$  required to produce a QSO luminosity.
Therefore, this merger can be regarded as a QSO  dominated by
activities induced by accretion power of SMBHs.
The epoch of maximum AR nearly coincides with that of maximum SFR, and
this coincidence can be seen in most of the present models.
These results  imply that (1) galaxy mergers with AGN activity can
contain starburst components, and (2) starburst components 
in the merger would not
be so easily detected owing to the  overwhelming light from the AGN.   
Since morphological transformation from spirals into an elliptical
is nearly finished at the epoch of the maximum AR in this model,
the merger can be regarded as a forming elliptical 
with starburst and AGN components.

As a natural result of the high ARs 
in the galactic nuclei,
the ratio of bolometric luminosity from the AGN 
($L_{\rm acc}$) and that from the starburst ($L_{\rm sb}$)
becomes very large in the final phase of galaxy merging.
Figure 3 shows that (1) $L_{\rm acc}/L_{\rm sb}$ becomes
higher as $L_{\rm acc}$ becomes higher, 
(2) it  becomes
more than 100 
at the epoch of maximum $L_{\rm acc}$,
and (3) it is higher during and after the coalescence
of two disk galaxies than before.
These results  suggest that the central starburst component
in a merger 
can  be more difficult to detect when the AR
of the merger becomes higher. 
They also suggest that  young elliptical galaxies formed
by major galaxy merging are more likely to show 
%by AGN-like spectra rather than HII region ones.
spectra with AGN features than HII region features.
We will discuss this point %later.
in \S 4.1.

A key factor in the evolution of these systems is the presence of
feedback from the AGN. If we construct a model that has gas consumption
by SMBHs, but without AGN feedback, we find that the maximum 
${\dot{m}}_{\rm sf} $
and ${\dot{m}}_{\rm acc}$  
are increased by factors of 2.8 and 113.0 respectively compared to
model M1 (which has the feedback present).
This model also shows a higher residual star formation rate
(${\dot{m}}_{\rm sf} \sim 10 {\rm M}_{\odot} {\rm yr}^{-1}$)
in a sporadic way
even after coalescence of two cores in galaxy merging.
It is therefore clear from these results that
(1) AGN feedback can suppress both  (i) gas fueling to SMBHs and
(ii) nuclear starbursts and (2) AGN feedback can strongly suppress
residual star formation after coalescence of two cores.
Springel et al. (2005b) have already pointed out that
AGN feedback can expel the remaining gas from merger remnants
to shut off star formation in their sophisticated models
of AGN feedback. 
The self-control of the growth of the SMBH by AGN feedback effects may
be important for better understanding the origin of the Magorrian
relation (eg. Magorrian et al 1998). However, such discussion is outside
the scope of this paper.

Figure 4 shows the time evolution of $M_{\rm SMBH}$ initially
within bulges of the two merging disks. 
Both SMBHs grow quickly by a significant factor owing to
efficient gas fuelling  to the central $10-100$ pc
and the resultant formation
of massive accretion disks around them when
star formation rates are quite high ($>10 {\rm M}_{\odot} {\rm yr}^{-1}$).
The difference in the growth rates shown in Figure 2 is due to the fact
that gas fuelling in the less inclined disk galaxy (i.e.\ galaxy 1)
is more efficient than in the more inclined one (i.e.\ galaxy 2). 
The mass of the forming elliptical is $\sim 3M_{\rm d}$ 
corresponding to $\sim 1.8 \times 10^{11} {\rm M}_{\odot}$ 
and the final combined mass of SMBH1 and SMBH2 is 
$7.7 \times 10^{8} {\rm M}_{\odot}$.
Therefore, the remnant elliptical of this model
shows $M_{\rm SMBH}/M_{\rm sph} = 4.3 \times 10^{-3}$,
where $M_{\rm sph}$ is the total mass of the elliptical.
This ratio of $M_{\rm SMBH}/M_{\rm sph}$ is reasonably
consistent  with
the observed value of 0.006 (Magorrian et al. 1998). 
Given the fact that $M_{\rm SMBH}/M_{\rm sph} = 2.0 \times 10^{-3}$
for the model with no growth of SMBHs,
the result shown in Figure 4 suggests that
the growth of SMBHs during merging is quite important for
elliptical galaxies formed by merging to show 
$M_{\rm SMBH}-M_{\rm sph}$ relation similar to 
the observed one (Magorrian et al. 1998). 

\begin{figure}
\psfig{file=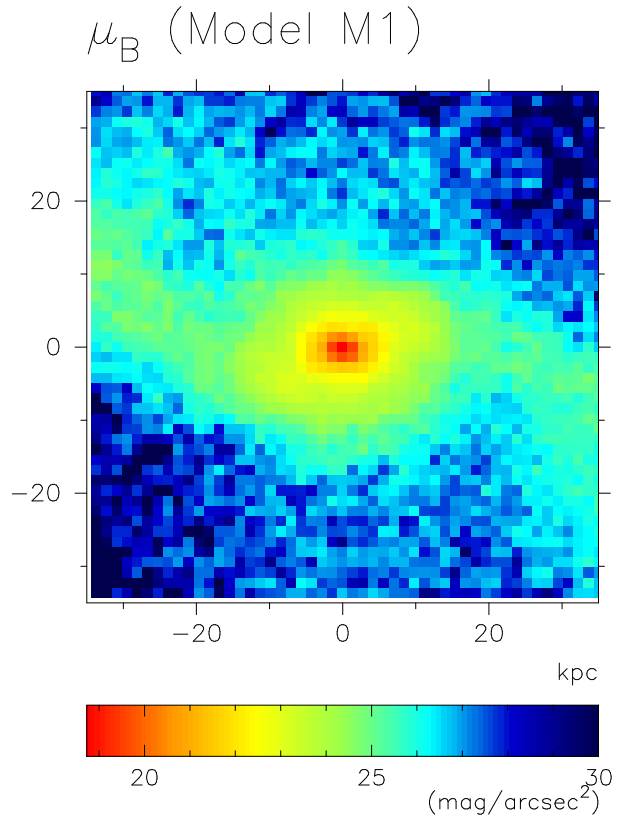,width=8.cm}
\caption{ 
$B-$band surface brightness (${\mu}_{\rm B}$) distribution 
at the epoch of the maximum AR in the standard model (M1).
}
\label{Figure. 6}
\end{figure}

\begin{figure}
\psfig{file=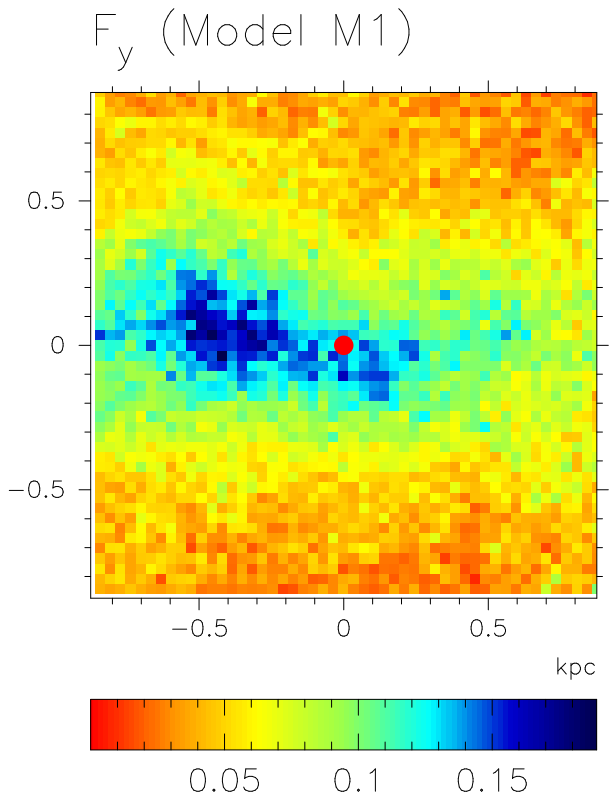,width=8.cm}
\caption{ 
The 2D distribution of the mass fraction
of new stars ($F_{\rm y}$) in the standard model (M1).
The location of the SMBH1 is indicated by a red filled circle.
Note that the SMBH is surrounded by young stellar populations.
}
\label{Figure. 7}
\end{figure}
Figure 5 describes the time evolution of projected mass densities
of new stars (${\Sigma}_{\rm ns}$))
and column densities (${\Sigma}_{\rm g}$),
which is a measure of the degree of dust extinction
for a given metallicity (e.g., Binney \& Merrifield 1998).
Figure 5 clearly shows
the significantly increased values of ${\Sigma}_{\rm ns}$
after $T \sim 1.8$ Gyr mean that SMBHs are surrounded by
compact poststarburst populations formed during galaxy merging.
The very large values of ${\Sigma}_{\rm g}$ around
$T=1.9$ Gyr mean
that SMBHs (thus AGN)  are heavily obscured  by metal-enriched
gas (thus dust).

Figure 6 shows the $B-$band surface brightness ($\mu_{\rm B}$) distribution of
the merger 
at the epoch of maximum AR.
Two disks are completely destroyed to form a spheroidal component 
by violent relaxation
until this epoch,
and only weak signs of tidal disturbance 
can be seen in its outer stellar halo.
Owing to the low surface brightness outer halo components
($\mu_{\rm B} > 27$  mag arcsec$^{-1}$),
this merger with a QSO-like activity can
be classified morphologically as an E if it is located at
high redshift with $z>1$ (e.g., Bekki et al. 1999 for
morphological properties of high $z$ starbursting mergers).
This result suggests that {\it some intermediate- and high-redshifts 
QSO host galaxies with apparently spheroidal morphologies
can be forming elliptical via dissipative major merger events.} 
This result accordingly appears to be consistent with 
an observational result (Floyd et al. 2004)
that QSO hosts at $z \sim 0.4$
are massive bulge-dominated  galaxies. 

Figure 7 shows the two dimensional (2D) distribution of $F_{\rm y}$
of the merger at the epoch of maximum AR,
where $F_{\rm y}$ is the mass fraction of new stars
(as a proportion of the total number) in the central regions of the merger.
This figure indicates that (1) the central SMBH (in the galaxy 1)
can be surrounded by young starburst or poststarburst components
in the central 1 kpc of the merger 
and (2) the location of the SMBH is 
however not necessarily coincident exactly with
the location  where most of the very young stellar components are formed
in the final phase of merging.
This difference in the locations of the SMBH and the starburst
region does not stay significant, as the two SMBHs
dynamically disperse the young compact starburst components when
they become closer to each other. 
Most of the present major merger models show the coexistence
of young starburst (or poststarburst) and AGN components
in the central 1 kpc of mergers
when morphological transformation is nearly completed.

We here stress that the derived coexistence of {\it moderately
strong starburst (${\dot{m}}_{\rm sf} \approx 30 {\rm M}_{\odot} {\rm yr}^{-1}$)
and QSO-like AGN (${\dot{m}}_{\rm acc} \approx 3 {\rm M}_{\odot} {\rm yr}^{-1}$)}
is due partly to gas consumption by the growth of accretion disks and SMBHs.
Our model with no gas accretion onto accretion disks and SMBHs (model M2) 
shows SFR of $\sim 640 {\rm M}_{\odot} {\rm yr}^{-1}$, which is significantly
higher than that of the standard model (See the 7th and 8th columns in the
table 1).
These comparative experiments indicate that
the presence of SMBHs that can swallow gas
and input feedback energy  can significantly influence
SFRs in galaxy mergers. 

Owing to rapid chemical enrichment from efficient star formation
during starburst phases of galaxy merging,
the stellar metallicities of stellar populations 
that are located within 100pc of SMBHs
at the maximum AR (i.e.\ QSO phases)
become as high as $2 {\rm Z}_{\odot}$,  where ${\rm Z}_{\odot}$
is the solar metallicity (=0.02).
The mean ages of new stars around SMBHs is 0.85 Gyr for SMBH1
and 0.71 Gyr for SMBH2 at the maximum AR of the merger.
Given the fact that Balmer absorption lines can become
strong (thus show ``E+A'' spectra)
$\sim 1$ Gyr after dusty starbursts (e.g., Bekki et al. 2001),
the above results strongly suggest that SMBHs in AGN-dominated ULIRGs
can be surrounded by metal-rich and young stellar populations
with strong Balmer absorption lines.
We discuss an evolutionary link between ULIRGs, QSOs, 
and E+A's later in \S 4.3.

\begin{figure}
\psfig{file=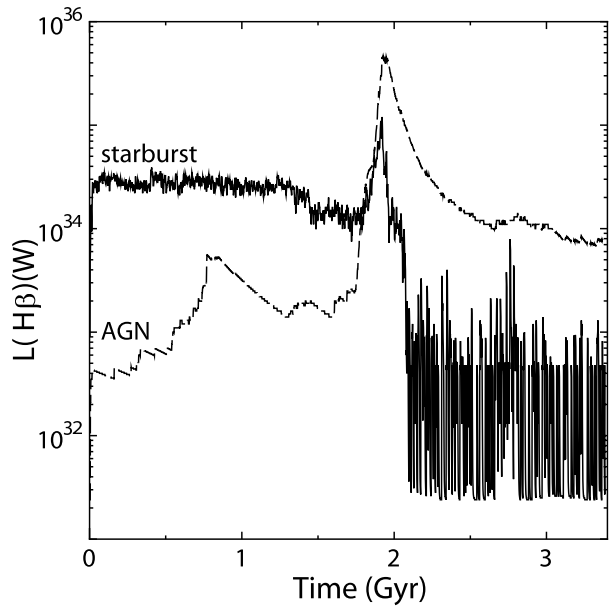,width=8.cm}
\caption{ 
Time evolution of  $L_{\rm H \beta}$ (in units of Watts) for the starburst
component (solid line) and the AGN component (dashed)
in the standard model. Note that the contribution from the AGN
strengthens significantly around T=1.9 Gyr, when the
merger becomes an AGN-dominated ULIRG.
}
\label{Figure. 8}
\end{figure}

\begin{figure}
\psfig{file=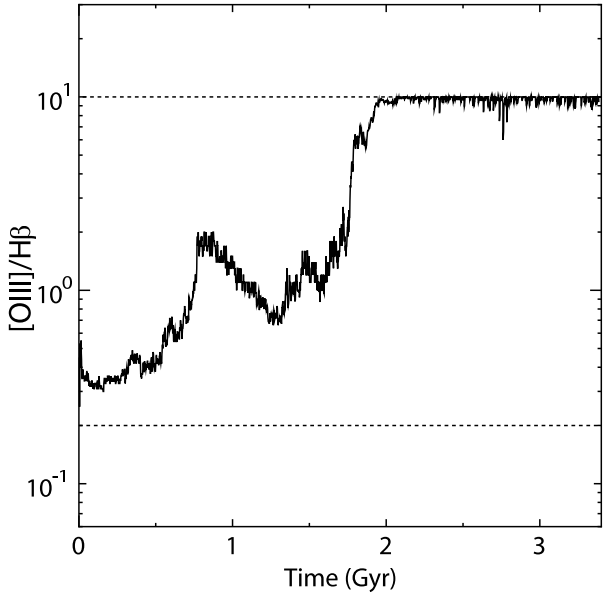,width=8.cm}
\caption{ 
Time evolution of the emission line ratio of 
$\oiii/{\rm H}\beta$ 
in the galaxy merger of the standard model.
Upper and lower (horizontal) dotted lines
represent typical values of starburst and AGN,
respectively.
Note that as galaxy merging proceeds,
the emission line ratio evolves from starburst-like one
into AGN-like one.
}
\label{Figure. 9}
\end{figure}

\begin{figure}
\psfig{file=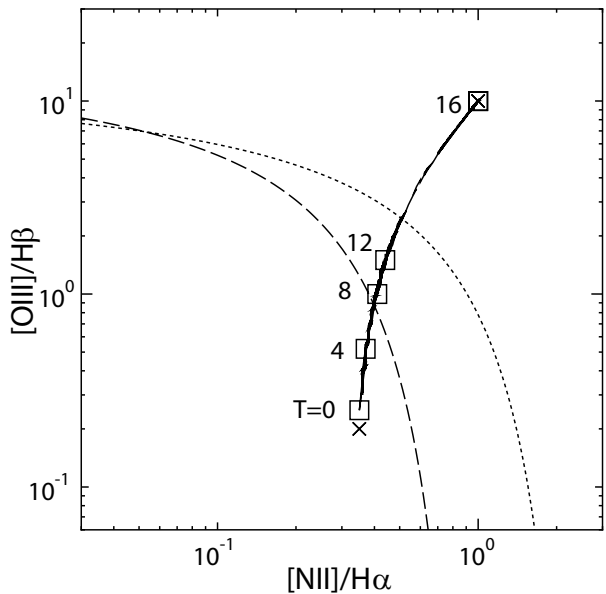,width=8.cm}
\caption{
Time evolution of the galaxy merger  in the standard model
on the $\oiii/{\rm H}\beta$-$\nii/{\rm H}\alpha$
diagram.
For clarity and comparison, only the results at selected epochs 
are described.
The results at 5 time steps (in simulation units) are indicated
by squares ($T=4$, 8, 12, 16 correspond to 0.56, 1.13,
1.69, and 2.26 Gyr, respectively). The lower and upper
crosses represent typical values for starbursts and AGN, respectively.
Dotted and dashed lines represent the  division
between starbursts and AGN by Kewley et al. (2001) and
Kauffmann et al. (2003b), respectively.
}
\label{Figure. 10}
\end{figure}

\begin{figure}
\psfig{file=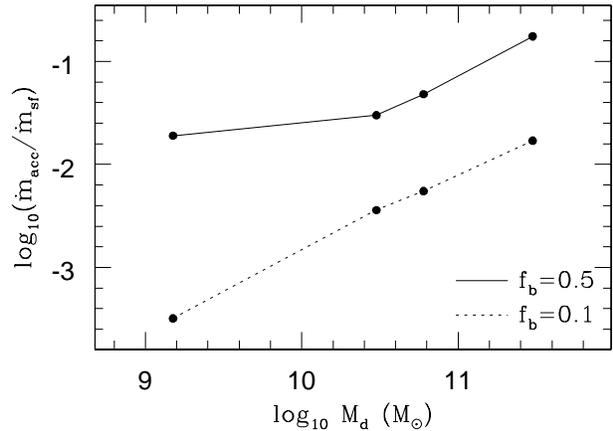,width=8.cm}
\caption{ 
Dependences of ${\dot{m}}_{\rm acc}/{\dot{m}}_{\rm sf}$
on the initial disk masses ($M_{\rm d}$) for the two sets
of models with $f_{\rm b}=0.5$ (solid) and $f_{\rm b}=0.1$ (dotted).
Both ${\dot{m}}_{\rm acc}$ and ${\dot{m}}_{\rm sf}$
are estimated at the epochs of their maximum values.
}
\label{Figure. 11}
\end{figure}

\begin{figure}
\psfig{file=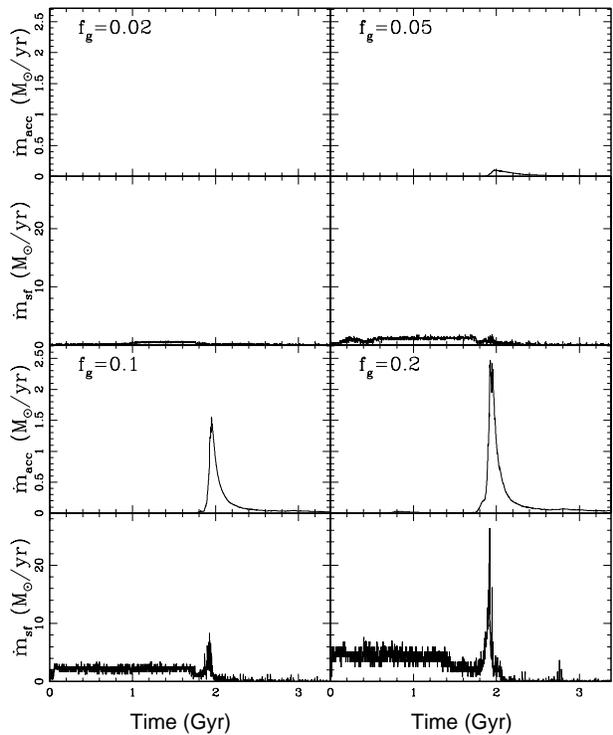,width=8.cm}
\caption{ 
Time evolution of ARs (${\dot{m}}_{\rm acc}$)
and SFRs (${\dot{m}}_{\rm sf}$) for four models with
different gas mass fraction: $f_{\rm g}=0.02$ (upper left),
$f_{\rm g}=0.05$ (upper right),$f_{\rm g}=0.1$ (lower left),
and $f_{\rm g}=0.2$ (lower right).
}
\label{Figure. 12}
\end{figure}

\begin{figure}
\psfig{file=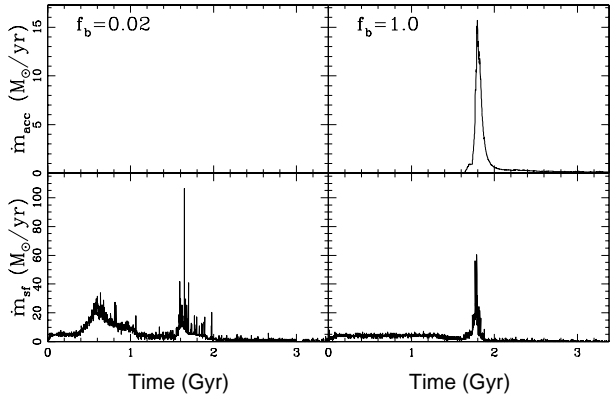,width=8.cm}
\caption{ 
Time evolution of ARs (${\dot{m}}_{\rm acc}$)
and SFRs (${\dot{m}}_{\rm sf}$) for two models with
different bulge-to-disk-ratios: $f_{\rm b}=0.02$ (left),
and $f_{\rm b}=1.0$ (right).
}
\label{Figure. 13}
\end{figure}

\begin{figure}
\psfig{file=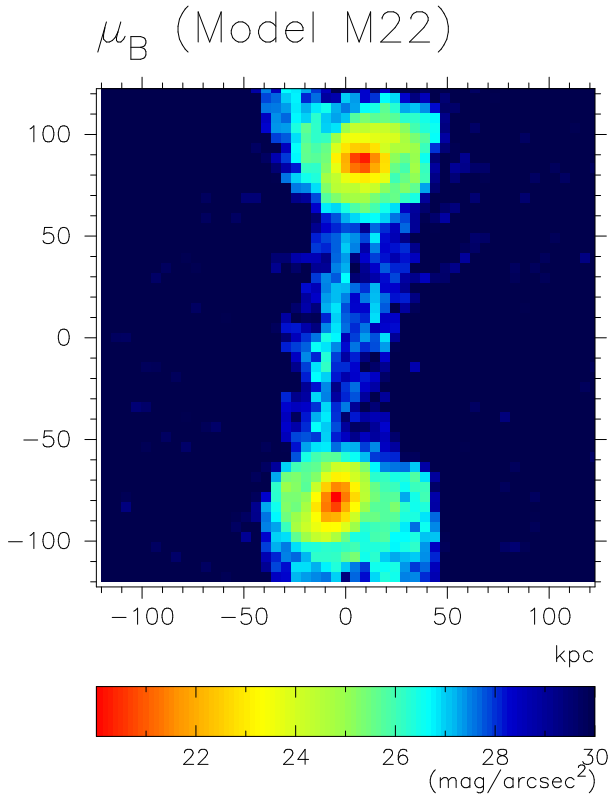,width=8.cm}
\caption{ 
The same as Figure 6 but for the tidal interaction model (M22).
}
\label{Figure. 14}
\end{figure}

\begin{figure}
\psfig{file=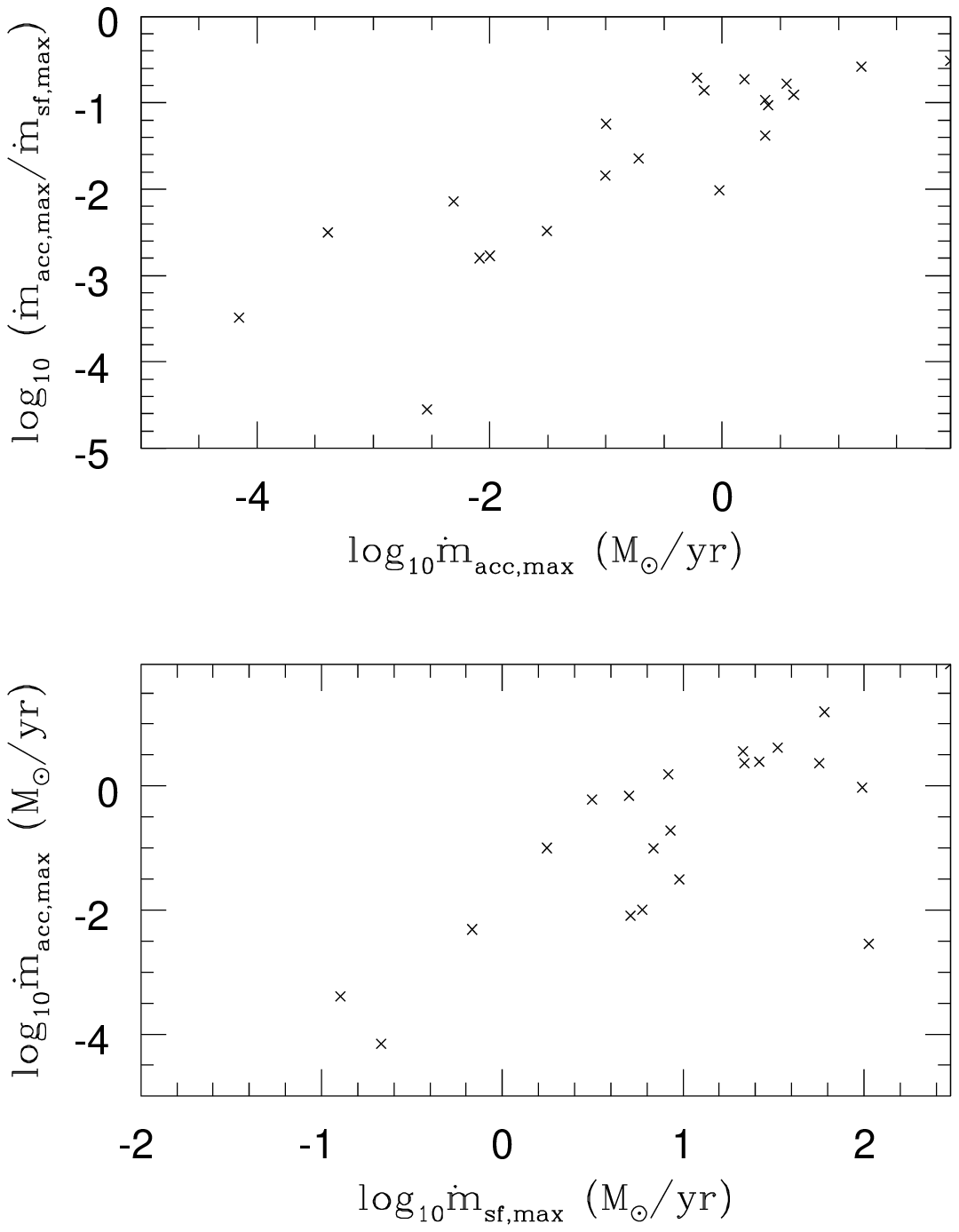,width=8.cm}
\caption{ 
The Dependence of the ratio of maximum AR (${\dot{m}}_{\rm acc,max}$)
to maximum SFR (${\dot{m}}_{\rm sf,max}$) on ${\dot{m}}_{\rm acc,max}$
(upper) and the dependence of ${\dot{m}}_{\rm acc,max}$ on
${\dot{m}}_{\rm sf,max}$ (lower) for the present 21 models. 
}
\label{Figure. 15}
\end{figure}

\subsubsection{Emission line properties}

Figure 8 shows time evolution of $L_{\rm H \beta}$
separately
for the starburst and AGN components.
Although $L_{\rm H \beta}$ of the starburst is larger than
that of the AGN in the early phases of galaxy merging ($T<1.8$ Gyr),
it becomes significantly smaller than that of the AGN
when gas fuelling to the SMBHs becomes efficient ($T \sim 1.9$ Gyr).
$L_{\rm H \beta}$ of the AGN component is always significantly
larger than that of the starburst one after the coalescence of
the two bulges: owing to very minor, sporadic star formation after
galaxy merging,  $L_{\rm H \beta}$ of the merger remnant
is dominated by the weak AGN component. 
Figure 9 clearly demonstrates that
the emission line ratio of \oiii/H$\beta$ 
of the merger changes from the value
typical for starbursts into that typical for AGN at $T \sim 1.9$ Gyr.
These results are consistent with the result (in Figure 3) that
$L_{\rm acc}/L_{\rm sb}$ becomes very high ($>100$)
at $T\sim 1.9$ Gyr.

Figure 10 shows the time evolution of the galaxy merger 
on the $\oiii/{\rm H}\beta$-$\nii/{\rm H}\alpha$
diagram, which is often used as a diagnostic for determining
whether spectral  properties of galaxies are dominated by starbursts 
or AGN
(e.g., Baldwin, Phillips \& Terlevich 1981; Veilleux \& Osterbrock 1987;
Kewley et al. 2001).
We also plot on the Figure two lines that demarcate the locations of
AGN- and starburst-dominated sources. The first
(dashed) line is the "extreme starburst classification line" from
Kewley et al (2001), a theoretically-derived upper limit for
starburst models. The second (dotted) line comes from Kauffman et al
(2003b), and is based on a large observational set
of data from the Sloan Digital Sky Survey. Starburst spectra should
reside below these lines, while AGN spectra should
be found to the upper right. 
It is clear from Figure 10 that the merger evolves from 
the middle of the starburst region 
%(where observed starburst galaxies are located)
toward the AGN-dominated region in the upper right.
%(where observed galaxies with AGN are located)
%on this emission line ratio diagram.

The results in Figures 8, 9 and 10 clearly show  that
there is an evolutionary link between starburst and AGN 
in terms of the spectral properties of the merger.
These results are consistent with the derived
SFRs and ARs (e.g., Figures 2 and 3), furthermore demonstrating the 
capability of our new codes in  correctly predicting
spectroscopic properties of galaxy mergers
with a coexistence of starbursts and AGN.
Full discussions on emission and absorption line properties
of  (including spectral lines other than those discussed above,
e.g., H$\gamma$ absorption line) of galaxy mergers
will be given in our
forthcoming papers (e.g., Bekki \& Shioya 2005).

\subsection{Parameter dependences}

Although the numerical results
on the coexistence of starbursts and AGN 
are similar 
for most of the merger models considered,
the magnitudes of SFRs and ARs 
depend on $M_{\rm d}$, $f_{\rm g}$, $f_{\rm b}$, $m_{2}$,
and orbital configuration of merging. 
We illustrate here the derived dependences and some physical correlations
between SFRs and ARs in galaxy mergers.

\subsubsection{$M_{\rm d}$}

Figure 11 shows how the relative importance of
starbursts and AGN in galactic nuclei of galaxy mergers depends
on the masses of the progenitor disks.
For two sets of models with different bulge-to-disk-ratio
($f_{\rm b}$ = 0.1 and 0.5), 
$\dot{m}_{\rm acc}/\dot{m}_{\rm sf}$
is larger for larger $M_{\rm d}$,
which  means that more massive  galaxy mergers 
are likely to be dominated by AGN rather than
by starbursts.
This is essentially because the interstellar gas that
is radially transferred from outer parts of mergers 
can be consumed more by SMBHs than by starbursts
owing to the larger initial masses of SMBHs 
(i.e.\  $M_{\rm SMBH} = 0.006f_{\rm b}M_{\rm d}$)
in more massive  mergers.   
We discuss this result later (\S 4) in the context
of the origin of ULIRGs.

\subsubsection{$f_{\rm g}$}

Figure 12 shows that maximum SFR and ARs in
galaxy mergers with larger  gas mass
fraction ($f_{\rm g}$) are both higher  
compared to those with smaller  gas mass fraction.
This is because the total amount of gas transfered to
the vicinity of SMBHs is larger for the mergers with
larger $f_{\rm g}$ owing to a larger amount of
gaseous dissipation in these.
Figure 12 also shows that mergers with $f_{\rm g} < 0.1$
exhibit ARs much smaller than the $0.7{\rm M}_{\odot} {\rm yr}^{-1}$
required for QSO activity.
These results suggest that $f_{\rm g}$ is one of key parameters
that determine whether galaxy mergers show QSO activity 
in their nuclei.
Irrespective of  $f_{\rm g}$,
the epoch of maximum SFR and that of maximum AR
are nearly coincide with each other,
which implies that coexistence of starbursts and AGN
can be quite common phenomena in galaxy mergers.

\subsubsection{$f_{\rm b}$}

Figure 13 describes the time evolution of SFRs and ARs for
the two extreme cases of mergers with $f_{\rm b}=0.02$
(nearly bulge-less spiral progenitor) and $f_{\rm b}=1.0$
(early-type M31-like spiral progenitor).
It is clear from this figure that maximum accretion
rates at the epoch of maximum SFRs are quite
different to each other in these two cases:
maximum ${\dot{m}}_{\rm acc}$ is $1.6 \times 10 {\rm M}_{\odot} {\rm yr}^{-1}$
for $f_{\rm b}=1.0$ 
and $3.0 \times 10^{-3} {\rm M}_{\odot} {\rm yr}^{-1}$ 
for $f_{\rm b}=0.02$.
This is due to the fact that the initial $M_{\rm SMBH}$ can determine
ARs in the present  model for the formation and the growth
of accretion disks. 
Figure 13 also shows that there is little difference in the maximum
SFR between the two models.
This result suggests that $f_{\rm b}$ is also a key parameter which
can determine whether galactic nuclei of mergers can be dominated by
starbursts or AGN.

The accretion radius ($R_{\rm B}$) within which gas clouds can be converted
into accretion disks is initially small for small SMBHs   
in the model (M19) with $f_{\rm b}=0.02$, owing to the adopted
assumption of $R_{\rm B} \propto M_{\rm SMBH}$.
Therefore, gas clouds have to lose a larger amount of angular momentum
(with respect to SMBHs) to reach $R_{\rm B}$ during merging.
As a result, gas is consumed by star formation rather than
by the growth of accretion disks around SMBHs.
Thus, the AR is significantly smaller compared with other major merger models
with bigger bulges. It should however be stressed here that 
if we relax the assumption of $R_{\rm B} \propto M_{\rm SMBH}$,
the AR in the model M19 can also reach high values.

Thus it should be stressed  that it depends on
the models of accretion radius ($R_{\rm B}$ 
dependent  on $M_{\rm SMBH}$ or not) 
whether initially small  MBHs 
(with masses of  $\sim 10^6 {\rm M}_{\odot}$ in the models with
small $f_{\rm b}$)
can grow to become SMBHs (with masses of $\sim 10^6 {\rm M}_{\odot}$).
Although both previous models (e.g., Springel et al. 2005b)
and the present one assume a relation between
$R_{\rm B}$ and $M_{\rm SMBH}$, it is not so clear (theoretically) whether
there really exists such a relation around SMBHs in galaxies. 
It would be therefore safe to say that future very 
high-resolution numerical simulations,
in which  a $R_{\rm B}-M_{\rm SMBH}$  relation is more 
self-consistently determined from sub-pc scale gas dynamics
around SMBHs, 
will provide a more robust prediction
on this matter.

\subsubsection{$m_{2}$}

Both maximum SFRs and ARs depend strongly on $m_{2}$
such that they are higher in mergers with
larger $m_{2}$
(See the 7th and 8th columns of the table 1 for the models M1, M13, and M14).
This is because a larger amount of interstellar gas
can be driven into the central regions of galaxy
mergers owing to stronger tidal disturbance 
and the resultant larger amount of shock dissipation
in mergers with larger $m_{2}$.
${\dot{m}}_{\rm acc}$ in minor (M13) and unequal-mass (M14)
mergers are well below $1{\rm M}_{\odot} {\rm yr}^{-1}$, so that
these merger remnants do not show QSO activity and thus may
well be identified as low luminosity AGN.
Since these minor and unequal-mass mergers ultimately
become S0s whereas major mergers can become Es (Bekki 1998),
S0s are more likely to show low luminosity AGN activity.
These results also imply that there can be a correlation
between AGN host morphological types (e.g, Es or S0s)
and nonthermal luminosities of AGN.

The HSB minor merger model (M15) shows a significantly high AR
(${\dot{m}}_{\rm acc} = 0.2{\rm M}_{\odot} {\rm yr}^{-1}$)
compared with the LSB counterpart M13
(${\dot{m}}_{\rm acc} = 0.01{\rm M}_{\odot} {\rm yr}^{-1}$), which suggests that
the compactness of the smaller galaxy in a minor merger
is a key factor for gas fuelling to the central SMBHs.
The reason for this high AR is that the smaller galaxy
of the HSB model can not be destroyed by the larger galaxy
until the final coalescence of the two galaxies 
so that it tidally disturbs the ISM of the larger galaxy
more strongly and for a longer time and thus triggers more
efficient gas fuelling.

\subsubsection{Orbital configurations}

No significant dependences of time evolution of SFRs and ARs on
orbital configurations are found for a given set of parameters
(See the 7th and 8th columns of
table 1 for the models, M1, M10, M11, and M12).
Typical SFRs and ARs are of the order of $10{\rm M}_{\odot} {\rm yr}^{-1}$
and $1{\rm M}_{\odot} {\rm yr}^{-1}$ respectively,
in these major merger models.
Given the derived $m_2$ dependences,
these results suggest major merging ($m_2 > 0.3$) is one of requisite
conditions for QSO formation. 

\subsubsection{Tidal interaction}

Both SFRs and ARs can be significantly enhanced  in tidal interaction
models, however the degree of the enhancement is much 
less remarkable compared with major merger models. 
Major merger models with bigger bulges (i.e.\ larger $f_{\rm b}$) 
can show larger ARs, whereas tidal interaction models with bigger
bulges do not show high ARs ($\sim 0.7{\rm M}_{\odot} {\rm yr}^{-1}$). 
This is because formation of strong stellar bars, which are the main drivers
for gas fuelling to starbursts and AGN, can not be formed in
the bigger bulge models (e.g., $f_{\rm b}=0.5$). 
%Therefore, only tidal interaction models with smaller bulges 
%and larger disk masses can show high ARs.
Thus, tidal interaction models show high ARs only when they involve
smaller bulges and larger disk masses.

Figure~14 shows the 2D distribution of
${\mu}_{\rm B}$ in the tidal interaction model M22, with 
$M_{\rm d} = 3.0 \times 10^{11} {\rm M}_{\odot}$,
$f_{\rm b} = 0.1$ ,
and $M_{\rm SMBH} = 0.012 M_{\rm b}$  at the epoch of its maximum
AR ($2.4{\rm M}_{\odot} {\rm yr}^{-1}$).
Although we have investigated several representative
tidal interaction models,
only this tidal model with more massive SMBHs shows a sufficiently 
high AR ($>0.7{\rm M}_{\odot} {\rm yr}^{-1}$) to become a
QSO.
It is clear from Figure~14 that
interacting galaxies with QSO activity can be identified
as two galaxies. This is quite different from major merger
models in which a merger with a QSO activity almost
always shows a single elliptical morphology. 
Given the very limited range of parameters for tidal
interaction models to show QSO activities,
these results imply that,
if they are formed by galaxy interaction and merging, 
(1) most of QSO hosts can be elliptical galaxies 
and (2) binary QSOs can be very rare. 

This is borne out by observational results. Mortlock et al. (1999)
found only 16 binary QSOs, and calculated an ``activation radius'' of
between 50 and 100~kpc (cf.\ Figure~14). Assuming these were formed in
a galaxy-galaxy collision, this implies that QSO formation occurs late
in the collision process. A more recent survey of binary QSOs from the
Sloan Digital Sky Survey and 2dF Quasar Redshift Survey (Hennawi et
al. 2005) found 218 quasar pairs with separations $<1 h^{-1} {\rm
Mpc}$, implying a binary fraction of $\sim1$ in 1000.

\subsubsection{Correlations between $\dot{m}_{\rm sf}$ and 
$\dot{m}_{\rm acc}$}
Figure 15 shows that (1) there is a weak yet positive correlation
between maximum SFRs (${\dot{m}}_{\rm sf,max}$)
and maximum ARs (${\dot{m}}_{\rm acc,max}$)
and (2) there is a clearer
correlation between ${\dot{m}}_{\rm acc,max}$
and ${\dot{m}}_{\rm acc,max}/{\dot{m}}_{\rm sf,max}$.
The above result (1) suggests that mergers with more pronounced
starburst activities are likely to show more pronounced
AGN ones.
The result (2) suggests that mergers with more pronounced AGN activities
are more likely to be dominated by AGN rather than starbursts.
It should be however stressed here that if we plot
${\dot{m}}_{\rm sf}$ and ${\dot{m}}_{\rm acc}$ from data
at every time step of all models  (including non-AGN and non-starburst
phases) in the same way
as shown in Figure 10,
the derived two correlations becomes rather weak. 
Thus the correlations can be held only for mergers with strong starbursts 
and AGN.

\section{Discussions}

\subsection{Relative importance
of starbursts and AGN}

Our numerical simulations 
have shown that galactic mass is a key factor in determining
whether a forming early-type galaxy is dominated by starbursts or AGN.
The present study has demonstrated that
the bolometric luminosity ratio, $L_{\rm acc}/L_{\rm sf}$, 
is larger for galaxy mergers with larger initial disk masses ($M_{\rm d}$).
An order of magnitude  estimation can allow us to understand this result. 
For global galactic star formation, 
we adopt the Schmidt law, 
in which  ${\dot{m}}_{\rm sf} \propto {{\mu}_{\rm g}}^{1.5}$,
where ${\mu}_{\rm g}$ is the 
surface gas density of a disk.
We also adopted the $M_{\rm d}-{\mu}_{\rm s}$ relation
(Kauffman et al. 2003b) in the present study.
Therefore $L_{\rm sf}$  can be approximated as: 
\begin{equation}
L_{\rm sf} \propto {\dot{m}}_{\rm sf} 
\propto {{\mu}_{\rm g}}^{1.5}  \propto {M_{\rm d}}^{0.75},
\end{equation}
for  ${\mu}_{\rm g} \propto {\mu}_{\rm s}$ in
the present models with radially constant gas mass fraction.
On the other hand,  $L_{\rm acc}$  can be approximated as:
\begin{equation}
L_{\rm acc} \propto {\dot{m}}_{\rm acc} 
\propto {\dot{m}}_{\rm Edd}  
\propto M_{\rm SMBH} \propto M_{\rm b} \propto M_{\rm d}
\end{equation}
for a given bulge-to-disk-ratio ($f_{\rm b}$).
Equation 16 and 17 lead us to derive the following 
relation:
\begin{equation}
\frac{L_{\rm acc}}{L_{\rm sf}}  \propto
\frac{{\dot{m}}_{\rm acc}}{{\dot{m}}_{\rm sf}} 
\propto {M_{\rm d}}^{0.25}.
\end{equation}
This relation suggests that more luminous forming early-type
galaxies via galaxy merging are likely to be dominated
by AGN rather than by starbursts, if merger progenitor disks
contain a sufficient amount of gas for fuelling.

Although the above analytically derived relation of 
$\frac{{\dot{m}}_{\rm acc}}{{\dot{m}}_{\rm sf}} 
\propto {M_{\rm d}}^{0.25}$ is 
qualitatively consistent with the simulations shown in in Figure 11,
it is significantly shallower than those derived in
the simulations 
($\frac{{\dot{m}}_{\rm acc}}{{\dot{m}}_{\rm sf}} 
\propto {M_{\rm d}}^{0.95}$ for $f_{\rm b}=0.1$
and $\frac{{\dot{m}}_{\rm acc}}{{\dot{m}}_{\rm sf}} 
\propto {M_{\rm d}}^{0.38}$ for $f_{\rm b}=0.5$; See Figure 11).
The origin of this difference might  well be closely associated
with the fact that suppression of star formation from
AGN feedback (which can enhance  the relative importance
of accretion-power-induced activity in galactic nuclei)
is not explicitly considered in the above analytical
arguments.

It is currently less feasible to prove the above mass dependence
based on the comparison between the simulation results and observations,
because most of previous observations focused on correlations between
nuclear activities and the Hubble morphological types 
(e.g., Mouri \& Taniguchi 2004).
It may well be an observationally difficult task
to estimate {\it separately} ${\dot{m}}_{\rm sf}$ and  ${\dot{m}}_{\rm acc}$
from emission line properties of galactic nuclei
for determining $L_{\rm acc}/L_{\rm sf}$.
We however suggest  that future statistical studies  on $L_{\rm acc}/L_{\rm sf}$
and its dependence on galactic masses are  worthwhile,
because they can prove an example of  mass-dependent evolution of galaxies.

\subsection{What powers ULIRGs ?}

It has been a longstanding, remarkable problem 
what dominate the luminosities of ULIRGs since 
many observational studies with different wavelengths
revealed possible evidences for both starbursts and AGN
in ULIRGs
(e.g., Sanders \& Mirabel 1996; Lutz et al. 1998;
Genzel et al. 1998; Tacconi et al. 2002).
Based on a mid-infrared spectroscopic survey of 15 ULIRGs
by {\rm ISO (Infrared Space Observatory}, 
Genzel et al. (1998) revealed that there is no obvious
trend for the AGN component to dominated in the most advanced
mergers.
Lutz et al (1998) investigated the ratio of the 7.7 $\mu$m PAH
(polycyclic aromatic hydrocarbon) emission feature to the local
continuum for 60 ULIRGs
and found that only about 15\% of ULIRGs at luminosities 
below $2\times 10^{12} L_{\odot}$ are powered by AGN.

Our simulations have demonstrated  that 
(1) initial galactic masses can be one of primarily important
parameters that determine the relative importance of starbursts
and AGN in galaxy mergers
and (2) more massive galaxy mergers  are likely to be dominated
by AGN.
Almost all ULIRGs show strongly disturbed morphological properties,
which are the most likely to be clear signs of past major merger events
(e.g., Sanders et al. 1988).
A logical conclusion of these theoretical and 
observational results  is that 
if ULIRGs are more massive 
then they are likely to be dominated by AGN.
Figure 11 suggests that 
galaxy mergers with their progenitor disk masses higher than
$\approx 10^{11} M_{\odot}$
and bigger bulges can become ULIRGs with AGN.
Recently Tacconi et al. (2002) have investigated structural and
kinematical parameters of ULIRGs and found that ULIRGs
are not so massive/bright  as giant ellipticals. 
This result, combined with our simulations, suggests  that
ULIRGs are, {\it on average}, dominated by starbursts
rather than AGN.
This suggestion is broadly consistent with the observational results by
Lutz et al. (1998) that about 80\% of ULIRGs are
found to be predominantly powered by starbursts.
It is however not so clear why galaxy mergers between less luminous
late-type galaxies are more likely to occur than those
between more luminous ones at lower redshifts.

\subsection{Evolutionary link between ULIRG,  QSOs, 
Q+A's,  and E+A's ?}

Although the relative importance of starbursts and AGN in ULIRGs
has been observationally suggested to be different between
different ULIRGs (e.g., Genzel et al. 1998; Lutz et al. 1998),
a significant fraction of ULIRGs have been  suggested to contain starburst
components (e.g., Farrah et al. 2003).
These observational results raise the following question: Are there
any evolutionary links between ULIRGs and  galaxies with ``E+A'' spectra
indicative of poststarburst populations (e.g., A-type stars) ?
This question may well be quite timely and important, given the 
fact that physical properties of E+A's are now being extensively
investigated for a large number of E+A samples derived by 
wide field surveys (e.g., Blake et al. 2004; Goto et al. 2003)
and by 8m-class ground telescopes with multi-object spectrograph
(e.g., Pracey et al. 2004).
Spectral signatures  of poststarburst stellar populations in some QSOs
(e.g., Canalizo \& Stockton 2000; 2001) and in some ULIRGs 
(e.g., Poggianti \& Wu 2000; Goto 2005)
imply that  there could be some close physical relationships between
ULIRGs, QSOs, and E+A's. In the following discussion, QSOs with
poststarburst spectra (i.e.\ strong Balmer absorption lines)
are referred to as ``Q+A's'' just for convenience.

The present simulations have demonstrated that SMBHs 
of galaxy mergers can be surrounded
by circumnuclear, compact,
and young poststarburst populations when ARs onto SMBHs are high.
This result implies that if the Balmer absorption lines are not
significantly diluted  by Balmer emission lines from AGNs,
spectral signatures of poststarburst  
populations can be detected in galaxy mergers.
Based on these numerical results, we suggest the following
two different evolutionary paths between mergers (Mer)
and ellipticals (Es). For SB-dominated ULIRGs that
are formed by merging either between less luminous disks or between
disks with smaller bulges, 
strong Balmer absorption lines can be detectable owing
to less significant dilution of the absorption lines
by emission lines 
from weaker AGN components. Therefore
the evolutionary path could be;

$Mer \Rightarrow ULIRGs  \Rightarrow E+A's \Rightarrow Es.$

For AGN-dominated ULIRGs that
are formed by merging between more massive disks with prominent bulges, 
the evolutionary path could depend on whether
the lifetimes of QSOs are shorter than the lifetimes of A-type stars.
The present simulations have shown that the lifetime of QSOs
(defined as the duration within which ${\dot{m}}_{\rm acc}$
is higher than 0.7 ${\rm M}_{\odot}$) is an order of $\sim 0.1$ Gyr
(See Hopkins et al. 2005 for possible luminosity dependent QSO lifetimes).
The dilution of the Balmer absorption lines by AGN emission
can be significant in the very  strong AGN phases 
so that the absorption lines can not be detected so easily.
The absorption lines might well be detectable 
either when intrinsic AGN luminosities become significantly smaller
or when AGN are observed as type II (i.e.\ viewed from the edge of
the surrounding dusty torus).
Therefore there can be the following evolutionary path:

$Mer \Rightarrow ULIRGs  \Rightarrow QSOs \Rightarrow (Q+A's) 
\Rightarrow E+A's \Rightarrow Es$.

Probably, QSOs that experienced stronger starbursts in 
the gas fuelling (thus growth) processes of SMBHs are more likely
to have detectable Balmer absorption lines.
It is also reasonable to claim that  strong Balmer absorption lines
are more likely to be detected in type 2 Seyfert than
in type 1 owing to the less amount of dilution of stellar
light by emission from hidden broad line regions of type 2 Seyfert
(with weaker emission lines due to dusty torus around AGN).
We plan to investigate this point more quantitatively
by numerical simulations
and compare the results 
with already existing observational results (e.g.,  
Kauffmann et al. 2003a;  Cid Fernandes et al. 2004).

As shown in the present chemodynamical study,
the metallicities of stellar populations around SMBHs
in galaxy mergers
become very high ($>2Z_{\odot}$) 
owing to rapid chemical enrichment associated
with starbursts during galaxy merging.
Furthermore, as a natural result of chemical evolution,
the stellar metallicities around SMBHs are higher in
the later phase of galaxy merging.
These results imply that 
stellar populations in later AGN phases (e.g., QSOs  and Liners)
are more likely to be more metal-rich than those in
earlier starburst phases for galaxy mergers with starbursts
and AGN.
Our previous chemodynamical simulations suggested that
the abundance ratio of [Mg/Fe] after strong starburst of galaxy mergers 
can be significantly larger than the solar value due to the dominant
contribution of Type II supernovae (Bekki \& Shioya 1999).
Therefore we suggest that {\it QSOs with younger poststarburst
populations can show large [Mg/Fe]  ratios if QSOs 
are evolved from ULIRGs formed by galaxy mergers.}

\subsection{Pair vs multiple mergers in ULIRGs formation.}

The present study has investigated mergers between two disk galaxies
and thereby demonstrated that ULIRGs can be formed in the very late phase
of galaxy merging when two galaxies nearly complete their merging. 
Thus the present model can be more relevant to ULIRGs with single cores:
The presence  of ULIRGs with multiple cores observed 
in some ULIRGs (e.g.,  Borne et al. 2000; Colina et al. 2001; 
Bushouse et al. 2002)
can not be simply explained by the present pair merger models 
(Taniguchi \& Shioya 1998). Previous numerical simulations showed
that (1) a compact group of galaxies can be transformed into an elliptical
galaxy through multiple merging of the group member galaxies 
(Barnes 1989; Weil \& Hernquest 1996),
(2) repetitive and multiple starbursts can be triggered by multiple
merging of disk galaxies (Bekki 2001),
and (3) the origin of metal-poor, hot gaseous halo 
of field giant ellipticals can be closely
associated with tidal stripping of metal-poor gas in multiple mergers
(Bekki 2001).
However, previously numerical studies did not  calculate the accretion
rates onto SMBHs and the SEDs in  multiple mergers so that 
they could not provide any theoretical predictions as to
(1) whether multiple mergers can become AGN-dominated ULIRGs or
starburst-dominated ones, 
(2) in what physical conditions starbursts and AGN can be obscured heavily
enough to become ULIRGs emitting almost all energy in infrared bands,
and (3) what is the dynamical fate of multiple SMBHs fallen into the
central regions of the remnants of multiple mergers.

Thus we plan to investigate the above questions based on
more sophisticated, higher-resolution simulations that allow
us to study both dynamical evolution of multiple SMBHs and gas accretion
onto the SMBHs. 
The results of these future simulations,
combined with those of the present study,
  will allow us to (1) investigate
what types of compact groups (e.g., spiral-rich groups)
can become starburst-dominated ULIRGs or AGN-dominated ones 
(or much less luminous infrared galaxies) 
in their conversion processes into field elliptical galaxies
via multiple galaxy merging
and (2) discuss statistics of the observed morphological properties
(e.g., single or multiple cores) of ULIRGs
(e.g., Murphy et al. 1996; Zheng et al. 1999; 
Borne et al. 1999; 2000; Colina et al. 2001;
Cui et al. 2001; Bushouse et al. 2002; Goto 2005).  
These simulations will also help us to understand physical relationships
between compact group of galaxies, 
multiple mergers, 
ULIRGs with hot gaseous halos (Xia et al. 2002; Huo et al. 2004),  
QSOs with companion galaxies
(Stockton \& Ridgway 1991; 
Disney et al. 1995; Hutchings \& Morris 1995;  Bahcall et al. 1997),
``fossil group'' with the central giant ellipticals (e.g., Ponman et al. 1994;
Jones et al. 2003).

\section{Conclusions}
We have numerically investigated both SFRs and ARs 
in forming ULIRGs via gas-rich galaxy merging 
in an   self-consistent way.
Dependences of the time evolution of SFRs and ARs on model parameters
are mainly investigated. 
We summarize our principle results as follows:

(1) ULIRGs powered  by AGN can be formed by major merging between
luminous, gas-rich disk galaxies with prominent bulges containing
SMBHs owing to the efficient gas fuelling 
(${\dot{m}}_{\rm acc} > 1 {\rm M}_{\odot}$ yr$^{-1}$) to the SMBHs.  
AGN in these ULIRGs can be  
surrounded by compact poststarburst stellar populations
(e.g., A-type stars).
These results suggest that ULIRGs and QSOs can show strong Balmer
absorption lines.

(2) ULIRGs powered by starbursts 
with ${\dot{m}}_{\rm sf} \sim 100 {\rm M}_{\odot}$ yr$^{-1}$
can be formed by 
merging between gas-rich disk galaxies with small bulges
having the bulge-to-disk-ratio ($f_{\rm b}$) as small as 0.1.   
As long as the accretion radii ($R_{\rm B}$) of SMBHs are  proportional
to the masses  of the SMBHs,
galaxy mergers with smaller bulges are more likely to become
starburst-dominated ULIRGs (i.e., they  can not show AGN activity
owing to a smaller amount of gas accretion onto the SMBHs).

(3) The relative importance of starbursts and AGN can depend
on physical properties of merger progenitor disks, such as
$f_{\rm b}$, gas mass fraction, and total masses.
For example, more massive  galaxy mergers are more likely
to become AGN-dominated ULIRGs.

(4) For most models,  major mergers can become
ULIRGs powered either by starbursts or by AGN, when the two bulges
finally merge.  
Interacting disk galaxies can become
ULIRGs with well separated  two  cores ($>$ 20kpc) 
at their pericenter
only when they are  very massive
and have  small bulges. 
These suggest that it is highly unlikely
for interacting/merging pair of galaxies
to become ULIRGs with double/multiple nuclei.
 We note, however, the results of Veilleux
et al (2002), who found that about 7\% of ULIRGs in their sample have
nuclear separations in excess of 20kpc. This may suggest that ULIRGs can be
formed via alternate routes to the major mergers examined herein.

(5) Irrespectively of models,
interacting/merging galaxies show the highest accretion rates
onto the central SMBHs and the resultant rapid growth of the SMBHs, when
their star formation rates are very high.

(6) ARs can become high ($1 {\rm M}_{\odot} {\rm yr}^{-1}$)
enough to show QSO-like activities 
($L_{\rm bol} \approx 10^{12} {\rm L}_{\odot}$)  
mostly in major mergers between massive disk galaxies with remarkable bulges.
ARs however can not reach the required rates for QSOs
(${\dot{m}}_{\rm acc} \approx 0.7{\rm M}_{\odot} {\rm yr}^{-1}$)
in minor and unequal-mass mergers that form S0s.
These results therefore imply that only forming elliptical 
via major mergers can show QSO-like activities
whereas forming S0s (or early-type spirals
with big bulges) via minor and unequal-mass merging show low luminosity
AGN (e.g., type 1/2 Seyfert).

(7) Maximum ARs  (${\dot{m}}_{\rm acc,max}$) can correlate
with maximum SFRs (${\dot{m}}_{\rm sf,max}$) in the sense
that galaxy mergers with higher ${\dot{m}}_{\rm sf,max}$
are likely to show higher ${\dot{m}}_{\rm acc,max}$.
This suggests that mergers and ULIRGs with more pronounced AGN activities
are likely to show stronger starburst components in their nuclei.
The correlations can be discussed in the context of
recent observational results (e.g., Goto 2005)
on correlations between infrared luminosities of ULIRGs,
star formation rates, and AGN luminosities (measured from [OIII] 
emission lines).

(8) The ratio of ${\dot{m}}_{\rm acc,max}$ to ${\dot{m}}_{\rm acc,sf}$
can correlate with ${\dot{m}}_{\rm acc,max}$ in the sense
that galaxy mergers with higher ${\dot{m}}_{\rm acc,max}$
are likely to show higher ${\dot{m}}_{\rm acc,max}/{\dot{m}}_{\rm sf,max}$.
This implies that merger  and ULIRGs with higher AGN 
(thus total) luminosities
are likely to be dominated by AGN rather than by starbursts.
This result can be  also consistent with recent results on
AGN fraction as a function of infrared luminosities of galaxies
(e.g., Goto 2005).

(9) There could be evolutionary links between ULIRGs, Q+A's, QSOs,
and E+A's. Galaxy mergers between less massive disks are more likely
to evolve from starburst-dominated
ULIRGs into E+As without experiencing QSO phases,
whereas those between more massive disks with prominent
bulges can evolve from AGN-dominated ULIRGs, to QSOs (and/or Q+A's),
and finally to E+A's, if the lifetimes of QSOs are as short as
$\sim 0.1$ Gyr. Removal of gas reservoir for star formation
via supernovae and AGN feedback
could be  essentially important for the above evolutionary links.

(10) Time evolution of emission line properties of galaxies
with starbursts and AGNs is investigated based on SFRs and ARs
derived from chemodynamical simulations.
For example, simulated mergers are demonstrated to evolve
from those  with smaller  \oiii/H$\beta$ (starburst-dominated) to those with
larger \oiii/H$\beta$ (AGB-dominated).
It is suggested that strong Balmer absorption lines
are more likely to be detected in type 2 Seyfert than
in type 1 owing to the less amount of dilution of stellar
light by emission from hidden broad line regions of type 2 Seyfert.
Direct comparison between the predicted spectrophotometric properties
of galaxy mergers with dusty starbursts and AGNs 
and the corresponding observations will be done
in our forthcoming papers.

%\acknowledgment
\section*{Acknowledgments}
We are  grateful to the referee for valuable comments,
which contribute to improve the present paper.
KB  acknowledges the financial support of the Australian Research 
Council throughout the course of this work.
The numerical simulations reported here were carried out on GRAPE
systems kindly made available by the Astronomical Data Analysis
Center (ADAC) at National Astronomical Observatory of Japan (NAOJ).

%\appendix
%\section[]{Large gaps in L\lowercase{y}${\balpha}$ forests\\* due to fluctuations in line distribution}

\end{document}